\documentclass[12pt,twoside,fleqn]{article}
\usepackage{latexsym}
\usepackage{epsf}
\usepackage{graphicx}

\renewcommand{\baselinestretch}{1.0}
\def\be{\begin{equation}}
\def\ee{\end{equation}}
\def\bea{\begin{eqnarray}}
\def\eea{\end{eqnarray}}
\def\nnb{\nonumber}
\def\bbuildrel#1_#2^#3{\mathrel{\mathop{\kern 0pt#1}\limits_{#2}^{#3}}}
\def\slash#1{\setbox0=\hbox{$#1$}#1\hskip-\wd0\dimen0=5pt\advance
       \dimen0 by-\ht0\advance\dimen0 by\dp0\lower0.5\dimen0\hbox
         to\wd0{\hss\sl/\/\hss}}

\newcommand{\f}{\frac}

\newcommand{\TT}{\rule[-2mm]{0mm}{7mm}}

\newcommand{\newsection}[1]{\section{#1}\setcounter{equation}{0}}

\newcommand{\hU}{{\hat U}}
\newcommand{\hJ}{{\hat J}}

\begin{document}
\begin{titlepage}

\begin{flushright}
  {\bf TUM-HEP-403/01\\       
       hep-ph/0102316
}\\[1cm]
\end{flushright}

\begin{center}

\setlength {\baselineskip}{0.3in} 
{\bf\Large Master Formulae for $\Delta F=2$ NLO-QCD Factors\\ 
in the Standard Model and Beyond}\\[2cm]

\setlength {\baselineskip}{0.2in}
{\large  Andrzej J. Buras, 
         Sebastian J{\"a}ger
         and J{\"o}rg Urban}\\[5mm]

{\it Physik Department, Technische Universit{\"a}t M{\"u}nchen,\\
     D-85748 Garching, Germany}\\[3mm]

{\bf Abstract}\\
\end{center} 
\setlength{\baselineskip}{0.2in}

We present analytic formulae for the QCD renormalization group factors
relating the Wilson coefficients $C_i(\mu_t)$ and $C_i(\mu)$, with
$\mu_t={\cal O}(m_t)$ and $\mu <\mu_t$, of the $\Delta F=2$ dimension 
six four-quark operators $Q_i$ in the
Standard Model and in all of its extensions. Analogous analytic
formulae for the QCD factors relating the matrix elements $\langle Q_i (2~
{\rm GeV})\rangle$ and $\langle Q_i (\mu_K)\rangle$ with $\mu_K<2$ GeV are
also presented. The formulae are given in the NDR scheme. 
The strongest renormalization-group effects are found for the operators with
the Dirac structures $(1 - \gamma_5) \otimes (1 + \gamma_5)$ and
$(1 - \gamma_5) \otimes (1 - \gamma_5)$. We calculate the matrix
elements $\langle \overline{K}^0|Q_i|K^0\rangle$ in the NDR scheme using the lattice
results in the LRI scheme. We give expressions for the mass
differences $\Delta M_K$ and $\Delta M_B$ and the CP-violating parameter
$\epsilon_K$ in terms of the non-perturbative parameters $B_i$ and the
Wilson coefficients $C_i(\mu_t)$. The latter summarize the dependence
on new physics contributions.

\end{titlepage}

\setlength{\baselineskip}{0.3in}

\newsection{Introduction}
\label{sec:intro}

Renormalization group short-distance QCD effects play an important
role in $K^0-\overline{K}^0$ and $B^0_{d,s}-\overline{B}^0_{d,s}$ mixing
within the Standard Model (SM) and its extensions \cite{BBL,AJBLH}. 
They can be
calculated by solving renormalization group equations that govern the
scale dependence of the Wilson coefficients $C_i(\mu)$ of the relevant
$\Delta F=2$ operators $Q_i$. The resulting effective weak Hamiltonian
reads
\be
H_{\rm eff}^{\Delta {\rm F}=2} = \frac{G_F^2}{16\pi^2}M_W^2 \sum_i V^i_{\rm CKM} 
C_i(\mu) Q_i~.
\ee
Here $G_F$ is the Fermi constant and $V^i_{\rm CKM}$ the
{Cabibbo-Kobayashi-Maskawa} (CKM) factor equal to $(V_{tb}^*V_{td})^2$ in
the case of $B^0_{d}-\overline{B}^0_{d}$ mixing in the SM. 
Beyond the SM other factors not proportional to CKM elements are generally present.
Using this Hamiltonian one can
calculate $\Delta F=2$ amplitudes, in particular the mass differences
 $\Delta M_K$ and $\Delta M_{d,s}$ in the  
$K^0-\overline{K}^0$ and $B^0_{d,s}-\overline{B}^0_{d,s}$ systems 
and the CP-violating  parameter $\varepsilon_K$.

Within the SM there is only one single operator
\be
Q_1^{\rm VLL} = \left(\bar{s}^\alpha \gamma_{\mu} P_L d^\alpha\right)
      \left(\bar{s}^\beta \gamma^{\mu} P_L d^\beta\right)
\label{smops}
\ee
relevant for $K^0-\overline K^0$ mixing, with analogous operators
for $B^0_{d,s}-\overline B^0_{d,s}$ mixing obtained from (\ref{smops})
through the appropriate change of flavours. Beyond the SM the full set
of dimension six operators contributing to $K^0-\overline{K}^0$ mixing
consists of 8 operators that can be split into 5 separate sectors
according to the chirality of the quark fields they contain. These
operators are listed in (\ref{normal}). Corresponding operators contributing 
to $B^0_{d,s}-\overline{B}^0_{d,s}$ mixing exist.

The general expression for $ C_i(\mu) $ is given by
\begin{equation}\label{CV}
 \vec C(\mu) = \hat U(\mu,\mu_t) \vec C(\mu_t)   
\end{equation}
where $ \vec C $ is a column vector built out of the $C_i $'s and
$\hat U(\mu,\mu_t)$ is the renormalization group matrix.
$\vec C(\mu_t)$, with $\mu_t={\cal O}(m_t)$, are the initial conditions 
which depend on the short distance physics at high energy scales.
In particular they depend on the top quark mass and the couplings 
and masses of new particles in extensions of the SM. We will later
briefly discuss the case of scales much higher than $m_t$. Otherwise
$\mu_t$ denotes a high energy scale in the range, say,
$M_W\leq\mu_t\leq 2 m_t$.

While the initial conditions $C_i(\mu_t)$ at the NLO level are known 
only in the SM \cite{{BJW90},{UKJS}} and in some of its extensions \cite{UKJS}, all the
ingredients are available to compute the NLO evolution matrix
$\hat{U}(\mu,\mu_t)$ for all possible extensions of the SM. Indeed, the
two-loop anomalous dimension matrix for all $\Delta F=2$
four-quark dimension six operators has been calculated in 
the regularization independent renormalization scheme (RI) 
in \cite{CET0} and in the NDR scheme  in \cite{BMU}. 
Together with the known one-loop anomalous dimension
matrix \cite{Gilman:1983ap,BMZ91} and the known $\beta$ function,
the evolution matrix can be
straightforwardly computed by means of the methods reviewed in
\cite{BBL,AJBLH}. 

The LO analytic expressions for $\hat U(\mu,\mu_t)$ can be found in
\cite{BMZ91}. For phenomenological applications it is  useful
to derive analogous expressions including NLO corrections. The first
step in this direction has been made in \cite{CET} where
$\hat U(\mu,\mu_s)$ with $\mu_s>m_t$ has been written as
\be     \label{two_step_ev}
\hat U(\mu,\mu_s)=\hat U(\mu,\mu_t)\hat U(\mu_t,\mu_s)
\ee
with $\hat U(\mu_t,\mu_s)$ given analytically in the Landau RI scheme
(LRI) but $\hat U(\mu,\mu_t)$
evaluated numerically for $\mu= 2$ GeV, $\mu_t=m_t$ and particular
values of $m_c$, $m_b$ and $\alpha_s$. The corresponding
results for $B^0_{d,s}-\overline{B}^0_{d,s}$ mixing have not been
presented in \cite{CET}.

It should be emphasized that NLO corrections are necessary for a
satisfactory matching of the Wilson coefficients to the matrix elements
obtained from lattice calculations.
Moreover as demonstrated in \cite{{BMZ91},{CET}} the inclusion of QCD
corrections at the LO and the NLO level is mandatory in order to place
reliable constraints on the parameters in the extensions of the SM, in
particular on the squark mass matrices in supersymmetric theories. 

The purpose of our paper is to present NLO analytic
formulae for the matrix $\hat U(\mu,\mu_t)$ relevant for 
$B^0_{d,s}-\overline B^0_{d,s}$ mixing $(\mu={\cal O}(m_b))$, and 
$K^0-\overline K^0$
mixing $(\mu={\cal O}(1-2~{\rm GeV}))$. These formulae when combined with
the initial conditions $\vec{C}(\mu_t)$ and the hadronic matrix elements 
$\langle\vec{Q}(\mu)\rangle$ will allow to calculate in the future 
the $\Delta F=2$ amplitudes for any extension of the SM.

The formulae given below for $\hat U(\mu, \mu_t)$ apply to the situation in
which the initial conditions for the Wilson coefficients are known at
$\mu_t={\cal O}(m_t)$ and the evolution down to scales $\mu<\mu_t$ is
performed in an effective theory with the top
quark and the heavy new particles integrated out. Whether the top
quark and the new particles have been integrated out at a single scale
$\mu_t$ or at different scales, say $\mu_t$, $\mu_{s_1}$, $\mu_{s_2}$
with $\mu_t<\mu_{s_1}<\mu_{s_2}$, is immaterial here. What matters are
the values of the Wilson coefficients at $\mu_t$ and not how they have
been evaluated from the contributions at scales higher than
$\mu_t$. On the other hand in the process of the evaluation of
$C_i(\mu_t)$ large logarithms $\log\mu_{s_1}/\mu_t$,
$\log\mu_{s_2}/\mu_{s_1}$ may appear. These logarithms have to
be resummed which results in new evolution functions $\hat
U(\mu_t,\mu_{s_1})$, $\hat U(\mu_{s_1},\mu_{s_2})$, etc. As discussed
in \cite{{BMZ91},{CET}} the structure of these matrices is model dependent and
consequently beyond the scope of the present paper. We will,
however, provide an analytic formula for the evolution  $\hat
U(\mu_t,\mu_s)$ with $\mu_t\ll\mu_s$ in an effective $f=6$ theory in
which only SM degrees of freedom  are present and all new particles
have been integrated out.

Now, the lattice results for the matrix elements
$\langle\vec{Q}(\mu)\rangle$ are usually given at $\mu=2$ GeV. In
what follows we will denote this scale by $\mu_L$. 
On the other hand large-$N$ approaches, the chiral quark model and any 
non-perturbative method in which the low-energy degrees of freedom 
are mesons provide these matrix elements at scales
$\mu_K\leq1$GeV. In our opinion it would be useful to have the matrix
elements obtained by means of different methods at a common 
``standard'' scale, which we will choose to be $\mu_L$ in the
following. This is achieved using the formula
\be
\langle\vec{Q} (\mu_L)\rangle = \hat{U}^T(\mu_K,\mu_L)
 \langle\vec{Q} (\mu_K)\rangle
\label{matelemevol}
\ee
where $\langle\vec{Q} (\mu_K)\rangle$ are the matrix elements
calculated for scales $\mu_K<\mu_L$ and $\hat{U}(\mu_K,\mu_L)$ is the
renormalization group evolution matrix. 
In our paper we provide analytic formulae for
$\hat{U}(\mu_K,\mu_L)$.

At this point we would like to stress that our paper is addressed first
of all to the practitioners of weak decays who do not want to get involved
with the details of NLO calculations but rather would like to use
the final QCD factors in phenomenological applications. On the other
hand it should also be useful to experts. Indeed, having explicit
analytic formulae, rather than numerical values, not only gives
the freedom to change input parameters but also makes
possible the
checking of a given calculation. In particular when multiplying the $\hat U$
matrices like in~(\ref{two_step_ev}) one easily generates
higher-order terms in $\alpha_s$ which really do not belong to NLO corrections.
While these corrections should be removed from NLO expressions, this is not
always done in the literature. Consequently, already at this stage unnecessary
discrepancies of the order of 5\% between calculations performed by
different groups may arise. These higher-order terms in $\alpha_s$ are
consistently removed in the present paper. We are aware of the fact
that some of the formulae presented below are rather long. Nevertheless
we believe that they should turn out to be useful in future phenomenological
applications.

The paper is organized as follows. In Section 2 we give the list of
the $\Delta F=2$ operators in question and establish our
notation. In Section 3 we give analytic formulae for the QCD
factors $\left[\eta_{ij}(\mu)\right]_a$ that represent the evolution matrix
$\hat{U}(\mu,\mu_t)$ in (\ref{CV}) in five different sectors, $a=({\rm
  VLL,\,LR,\,SLL,\,VRR,\,SRR)}$, in the leading order (LO) and the
next-to-leading (NLO) approximation in the NDR scheme. In  Section
4 we give the analogous formulae for the QCD factors
$\left[\rho_{ij}(\mu_K)\right]_a$ which represent the evolution matrix
$\hat{U}(\mu_K,\mu_L)$ in (\ref{matelemevol}). In Section 5 we provide
numerical results for $\left[\eta_{ij}(\mu)\right]_a$ and
$\left[\rho_{ij}(\mu)\right]_a$  in the NDR
scheme. In section 6 we discuss the transformation rules for obtaining the
corresponding results in other renormalization schemes and we present
the relation between the QCD factors calculated here and the QCD factors
$\eta_B$ and $\eta_2$ used in phenomenological applications.
In Section 7 we calculate the matrix
elements $\langle \overline{K}^0|Q_i|K^0\rangle$ in the NDR scheme using the lattice
results in the LRI scheme \cite{{CET},{latt_lit}}. We give general expressions for the mass
differences $\Delta M_K$ and $\Delta M_B$ and the CP-violating parameter
$\epsilon_K$ in terms of the non-perturbative parameters $B^a_i$ and the
Wilson coefficients $C_i(\mu_t)$. We conclude in Section 8. For completeness we list in
appendix~\ref{app_adm} the one-loop and two-loop anomalous dimension
matrices that we have used in our paper. Appendix~\ref{app_higher_orders}
gives the general formulae for the $\hat U$ matrices which have been used
to obtain the analytic formulae of sections~\ref{sec:eta} and~\ref{sec:rho}.
Finally in Appendix C we give analytic formulae for the evolution
matrix $\hat U (\mu_t, \mu_s)$.

\newsection{Basic Formulae}
\label{sec:basic}

For definiteness, we will give explicit expressions for the operators
responsible for the $K^0-\overline K^0$ mixing. The operators
belonging to the VLL, LR and SLL sectors read
\bea 
Q_1^{\rm VLL} &=& (\bar{s}^{\alpha} \gamma_{\mu}    P_L d^{\alpha})
                  (\bar{s}^{ \beta} \gamma^{\mu}    P_L d^{ \beta}),
\nnb\\[4mm] 
Q_1^{\rm LR} &=&  (\bar{s}^{\alpha} \gamma_{\mu}    P_L d^{\alpha})
                  (\bar{s}^{ \beta} \gamma^{\mu}    P_R d^{ \beta}),
\nnb\\
Q_2^{\rm LR} &=&  (\bar{s}^{\alpha}                 P_L d^{\alpha})
                  (\bar{s}^{ \beta}                 P_R d^{ \beta}),
\nnb\\[4mm]
Q_1^{\rm SLL} &=& (\bar{s}^{\alpha}                 P_L d^{\alpha})
                  (\bar{s}^{ \beta}                 P_L d^{ \beta}),
\nnb\\
Q_2^{\rm SLL} &=& (\bar{s}^{\alpha} \sigma_{\mu\nu} P_L d^{\alpha})
                  (\bar{s}^{ \beta} \sigma^{\mu\nu} P_L d^{ \beta}),
\label{normal}
\eea
where $\alpha,\,\beta$ are colour indices, $\sigma_{\mu\nu} = 
\f{1}{2} [\gamma_{\mu}, \gamma_{\nu}]$ and
$P_{L,R} =\f{1}{2} (1\mp \gamma_5)$. The operators belonging to the
two remaining sectors (VRR and SRR) are obtained from $Q_1^{\rm VLL}$ and
$Q_i^{\rm SLL}$ by interchanging $P_L$ and $P_R$. Since QCD preserves
chirality, there is no mixing between different sectors.  Moreover,
the anomalous dimension matrices and the evolution matrices in the VRR and SRR sectors are the same 
as in the VLL and SLL sectors, respectively.  Therefore, in the following, 
we shall consider only the VLL, LR and SLL sectors. However, one should
remember that the initial conditions $C_i(\mu_t)$ are generally changed
when $P_L$ and $P_R$ are interchanged.
The operators in the
case of $B^0_{d}-\overline{B}^0_{d}$ mixing are obtained from
(\ref{normal}) through the replacement $s\rightarrow
b$. Performing the subsequent replacement $d\rightarrow s$ gives the
operators contributing to $B^0_{s}-\overline B^0_{s}$ mixing. The
one-loop and two-loop anomalous dimension matrices of the operators~(\ref{normal})
are given in appendix~\ref{app_adm}.

Restricting the discussion to the VLL, LR and SLL sectors $\hat{U}
(\mu_1,\mu_2)$ takes the following form
\be
\hat{U} (\mu_1,\mu_2) =
\left(\begin{array}{ccc}
    \left[\eta (\mu_1,\mu_2)\right]_{\rm VLL} &0 & 0 \\
    0 &\left[\hat{\eta}(\mu_1,\mu_2)\right]_{\rm LR} &0 \\
    0 & 0 &\left[\hat{\eta}(\mu_1,\mu_2)\right]_{\rm SLL}
      \end{array}\right)
\ee
where $\left[\hat{\eta}(\mu_1,\mu_2)\right]_{\rm LR}$ and
$\left[\hat{\eta}(\mu_1,\mu_2)\right]_{\rm SLL}$ are $2\times 2$
matrices and $\mu_1<\mu_2$. 
In what follows we will use a short-hand notation,
denoting the QCD factors
representing $\hat{U}(\mu,\mu_t)$ and $\hat{U}(\mu_K,\mu_L)$ by
\bea
\label{etarho1}
\left[\hat\eta(\mu, \mu_t)\right]_a & \equiv &
\left[\hat\eta(\mu)\right]_a = \left[\hat\eta^{(0)}(\mu)\right]_a+
\frac{\alpha^{(f)}_s(\mu)}{4 \pi} \left[\hat\eta^{(1)}(\mu)\right]_a,\\
\label{etarho2}
\left[\hat\rho(\mu_K, \mu_L)\right]_a & \equiv &
        \left[\hat\rho(\mu_K)\right]_a = 
                         \left[\hat\rho^{(0)}(\mu_K)\right]_a+
\frac{\alpha^{(3)}_s(\mu_K)}{4 \pi}
        \left[\hat\rho^{(1)}(\mu_K)\right]_a,
\eea
respectively. That is, we will suppress the high-energy scale $\mu_t$
in the argument of the $\eta$-factors. Similarly, we will suppress the
``lattice scale'' $\mu_L$ in the argument of the $\rho$-factors.
Using this notation we have for
instance
\be
C_1^{\rm VLL}(\mu_b)=\left[\eta (\mu_b)\right]_{\rm VLL} 
                        C_1^{\rm VLL}(\mu_t),
\ee
\be
\left(\begin{array}{c}
C_1^{\rm LR}(\mu_b)   \\
C_2^{\rm LR}(\mu_b)
\end{array}\right)
=
\left(\begin{array}{cc}
\left[\eta_{11} (\mu_b)\right]_{\rm LR} &
\left[\eta_{12} (\mu_b)\right]_{\rm LR} \\
\left[\eta_{21} (\mu_b)\right]_{\rm LR} &
\left[\eta_{22} (\mu_b)\right]_{\rm LR}
\end{array}\right)
\left(\begin{array}{c}
C_1^{\rm LR}(\mu_t)   \\
C_2^{\rm LR}(\mu_t)
\end{array}\right),
\ee
\be
\langle Q_1^{\rm VLL}(\mu_L)\rangle =\left[\rho(\mu_K)\right]_{\rm VLL} 
                        \langle Q_1^{\rm VLL}(\mu_K)\rangle,
\ee
\be
\left(\begin{array}{c}
\langle Q_1^{\rm LR}(\mu_L)\rangle   \\
\langle Q_2^{\rm LR}(\mu_L)\rangle
\end{array}\right)
=
\left(\begin{array}{cc}
\left[\rho_{11} (\mu_K)\right]_{\rm LR} &
\left[\rho_{21} (\mu_K)\right]_{\rm LR} \\
\left[\rho_{12} (\mu_K)\right]_{\rm LR} &
\left[\rho_{22} (\mu_K)\right]_{\rm LR}
\end{array}\right)
\left(\begin{array}{c}
\langle Q_1^{\rm LR}(\mu_K)\rangle   \\
\langle Q_2^{\rm LR}(\mu_K)\rangle
\end{array}\right)
\label{rhot}
\ee
and analogous formulae for the SLL sector. Note that in accordance 
with (\ref{matelemevol}), the transpose of 
$\left[\hat\rho(\mu_K)\right]_{\rm LR}$
enters the transformation (\ref{rhot}).

In Section 3 we will give analytic formulae
for the LO factors $\left[\eta^{(0)}(\mu)\right]_{\rm VLL}$,
$\left[\eta_{ij}^{(0)}(\mu)\right]_{\rm LR}$,
$\left[\eta_{ij}^{(0)}(\mu)\right]_{\rm SLL}$ and the NLO factors
$\left[\eta^{(1)}(\mu)\right]_{\rm VLL}$,  
$\left[\eta_{ij}^{(1)}(\mu)\right]_{\rm LR}$,
$\left[\eta_{ij}^{(1)}(\mu)\right]_{\rm SLL}$. The corresponding
expressions for the $\rho$-factors are given in Section
4. $\alpha_s^{(f)} (\mu)$ is the QCD coupling constant in an
effective theory with $f$ flavours: $f=5$ for $\mu_b<\mu<\mu_t$, $f=4$
for $\mu_c<\mu<\mu_b$, and  $f=3$ for $\mu<\mu_c$, where  $\mu_t={\cal
  O}(m_t)$,  $\mu_b={\cal O}(m_b)$ and $\mu_c={\cal O}(m_c)$. We impose
the continuity relations
\be
\alpha_s^{(3)}(\mu_c)= \alpha_s^{(4)}(\mu_c),\qquad
\alpha_s^{(4)}(\mu_b)= \alpha_s^{(5)}(\mu_b)~.
\ee

The general expression for $\alpha_s^{(f)}$ reads
\begin{equation}\label{QCDC}
{{\alpha^{(f)}_s(\mu)}\over {4\pi}} =
{{1}\over{\beta_0 \ln\left(\mu^2/{\Lambda^{(f)}_{\overline{MS}}}^2\right)}}
- {{\beta_1}\over{\beta^3_0}} {{\ln \ln \left(\mu^2/{\Lambda^{(f)}_{\overline{MS}}}^2\right)}
\over{\ln^2\left(\mu^2/{\Lambda^{(f)}_{\overline{MS}}}^2\right)}}
\end{equation} 
with 
\be
\beta_0=11-\frac{2}{3}f,\qquad
\beta_1=102-\frac{38}{3} f.
\ee
$\Lambda_{\overline{MS}}^{(f)}$ is the QCD scale parameter in
a theory with $f$ quark flavours ~\cite{BBDM}. 
The existing analyses of high energy processes give  
$\alpha^{(5)}_{\overline{MS}}(M_Z)=0.118\pm 0.003$ \cite{Bethke} or 
equivalently $\Lambda^{(5)}_{\overline{MS}}=
(226\begin{array}{c}+41\\-36\end{array})$MeV.

The evolution matrix, $ \hat U(\mu,\mu_t) $,
is given as follows:
\begin{equation}\label{UM}
 \hat U(\mu,\mu_t) = T_g \exp \left[ 
   \int_{g(\mu_t)}^{g(\mu)}{dg' \frac{\hat\gamma^T(g')}{\beta(g')}}\right] 
\end{equation}
with $g$ denoting the QCD effective coupling constant and $T_g$ an
ordering operation defined in \cite{BBL}. $ \beta(g) $
governs the evolution of $g$ and $ \hat\gamma $ is the anomalous dimension
matrix of the operators involved. 

We also have 
\be
\label{UUU}
\hat U(\mu_K,\mu_t) = \hat U(\mu_K,\mu_c) \hat U(\mu_c,\mu_b) \hat
U(\mu_b,\mu_t)~
\ee
with the three factors on the r.h.s. evaluated in $f=3$, $f=4$ and
$f=5$ effective theories, respectively. Now,
\be
\label{UUU1}
\hat U(\mu_L,\mu_t) =  \hat U(\mu_L,\mu_b) \hat U(\mu_b,\mu_t), \qquad
\hat U(\mu_K,\mu_t) = \hat U(\mu_K,\mu_L) \hat U(\mu_L,\mu_t).
\ee
This means that knowing $\eta_{ij}(\mu_L)$ and $\rho_{ij}(\mu_K)$
allows to calculate $\eta_{ij}(\mu_K)$.

 Keeping the first two terms in the expansions of
 $\hat\gamma(g)$ and $\beta(g)$ in powers of $g$
\begin{equation} \label{beta_gamma}
\hat\gamma (g) = \hat\gamma^{(0)} \frac{\alpha_s}{4\pi} + \hat\gamma^{(1)}
\left(\frac{\alpha_s}{4\pi}\right)^2, 
\quad\quad
 \beta (g) = - \beta_0 \frac{g^3}{16\pi^2} - \beta_1 
\frac{g^5}{(16\pi^2)^2},
\end{equation}
inserting these expansions into (\ref{UM}) and (\ref{UUU}) and expanding 
in $\alpha_s$ 
one can calculate the $\eta$- and $\rho$-factors defined in
(\ref{etarho1}) and (\ref{etarho2}), respectively. To this end, one has to
remove terms ${\cal O}(\alpha_s^2)$ and higher-order terms. We discuss
this point in appendix~\ref{app_higher_orders} where the expansions of
$\hat U(\mu, \mu_t)$ for $\mu=\mu_b$, $\mu=\mu_L$ and $\mu=\mu_K$ are given.

\newsection{$\eta$-Factors in the NDR Scheme}
\label{sec:eta}

\subsection{$\eta$-Factors for $B^0_{d,s}-\overline{B}^0_{d,s}$
  mixing}

{\underline{VLL-Sector}}
\bea
\left[\eta^{(0)} (\mu_b)\right]_{\rm VLL} &=& \eta_5^{6/23},    \\
\left[\eta^{(1)} (\mu_b)\right]_{\rm VLL} &=&
                                1.6273 (1-\eta_5) \eta_5^{6/23}.
\eea
{\underline{LR-Sector}}
\bea
\left[\eta^{(0)}_{11} (\mu_b)\right]_{\rm LR} &=& \eta_5^{3/23},   \\
\left[\eta^{(0)}_{12} (\mu_b)\right]_{\rm LR} &=& 0,   \\
\left[\eta^{(0)}_{21} (\mu_b)\right]_{\rm LR} &=&
                        \frac23(\eta_5^{3/23} -\eta_5^{-24/23}),   \\
\left[\eta^{(0)}_{22} (\mu_b)\right]_{\rm LR} &=& \eta_5^{-24/23},\\
\left[\eta^{(1)}_{11}(\mu_b)\right]_{\rm LR}&=&
0.9250\,{\eta_{5}}^{ -{24/23}} + 
{\eta_{5}}^{ {3/23}}\,\left( -2.0994 + 1.1744\,\eta_{5} \right), \\
\left[\eta^{(1)}_{12}(\mu_b)\right]_{\rm LR}&=&
1.3875\,\left({\eta_{5}}^{ {26/23}} -\,{\eta_{5}}^{ -{24/23}}\right),\\
\left[\eta^{(1)}_{21}(\mu_b)\right]_{\rm LR}&=&
\left( -11.7329 + 0.7829\,\eta_{5} \right) \,{\eta_{5}}^{ {3/23}}+ 
{\eta_{5}}^{ -{24/23}}\,\left( -5.3048 + 16.2548\,\eta_{5} \right), \\
\left[\eta^{(1)}_{22}(\mu_b)\right]_{\rm LR}&=&
\left( 7.9572 - 8.8822\,\eta_{5} \right) \,{\eta_{5}}^{ -{24/23}}+ 
0.9250\,{\eta_{5}}^{{26/23}}.
\eea
{\underline{SLL-Sector}}
\bea
\left[\eta^{(0)}_{11} (\mu_b)\right]_{\rm SLL} &=&
                        1.0153 \eta_5^{-0.6315} - 0.0153 \eta_5^{0.7184},  \\
\left[\eta^{(0)}_{12} (\mu_b)\right]_{\rm SLL} &=&
                        1.9325 (\eta_5^{-0.6315} - \eta_5^{0.7184}), \\
\left[\eta^{(0)}_{21} (\mu_b)\right]_{\rm SLL} &=&
                        0.0081 (\eta_5^{0.7184} - \eta_5^{-0.6315}),    \\
\left[\eta^{(0)}_{22} (\mu_b)\right]_{\rm SLL} &=&
                        1.0153 \eta_5^{0.7184} - 0.0153 \eta_5^{-0.6315},    \\
\left[\eta^{(1)}_{11}(\mu_b)\right]_{\rm SLL}&=&
\left( 4.8177 - 5.2272\,\eta_{5}\right) \,{\eta_{5}}^{-0.6315} + 
\left( 0.3371 + 0.0724\,\eta_{5} \right) \,{\eta_{5}}^{0.7184},\\
\left[\eta^{(1)}_{12}(\mu_b)\right]_{\rm SLL}&=&
\left( 9.1696 - 38.8778\,\eta_{5} \right) \,{\eta_{5}}^{-0.6315} + 
\left( 42.5021 - 12.7939\,\eta_{5} \right) \,{\eta_{5}}^{0.7184},\\
\left[\eta^{(1)}_{21}(\mu_b)\right]_{\rm SLL}&=&
\left( 0.0531 + 0.0415\,\eta_{5}\right) \,{\eta_{5}}^{-0.6315} - 
\left( 0.0566 + 0.0380\,\eta_{5} \right) \,{\eta_{5}}^{0.7184},\\
\left[\eta^{(1)}_{22}(\mu_b)\right]_{\rm SLL}&=&
\left( 0.1011 + 0.3083\,\eta_{5}\right) \,{\eta_{5}}^{-0.6315} + 
{\eta_{5}}^{0.7184}\,\left( -7.1314 + 6.7219\,\eta_{5} \right), 
\eea
where 
\be
\label{eTa5}
\eta_5\equiv\frac{\alpha_s^{(5)}(\mu_t)}{\alpha_s^{(5)}(\mu_b)}.
\ee
\subsection{$\eta$-Factors for $K^0-\overline{K}^0$ mixing with $\mu=\mu_L$}
\label{sec_eta_at_muL}
These factors are relevant for $K^0-\overline{K}^0$ mixing but can
also be used in $D^0-\overline{D}^0$ mixing. They can be
expressed in terms of $\eta_5$ defined in (\ref{eTa5}) and 
\be
\label{eTa4}
\eta_4\equiv\frac{\alpha_s^{(4)}(\mu_b)}{\alpha_s^{(4)}(\mu_L)}.
\ee
{\underline{VLL-Sector}}
\bea
\left[\eta^{(0)} (\mu_L)\right]_{\rm VLL} &=&
                        \eta_4^{6/25} \eta_5^{6/23}, \\
\left[\eta^{(1)} (\mu_L)\right]_{\rm VLL} &=&
\eta_4^{6/25} \eta_5^{6/23} (1.7917-0.1644 \eta_4 -1.6273 \eta_4 \eta_5).
\eea
{\underline{LR-Sector}}
\bea
\left[\eta^{(0)}_{11} (\mu_L)\right]_{\rm LR} &=&
                        \eta_4^{3/25} \eta_5^{3/23},   \\
\left[\eta^{(0)}_{12} (\mu_L)\right]_{\rm LR} &=& 0,   \\
\left[\eta^{(0)}_{21} (\mu_L)\right]_{\rm LR} &=& 
                        \frac23 (\eta_4^{3/25} \eta_5^{3/23} -
                                \eta_4^{-24/25} \eta_5^{-24/23}),   \\
\left[\eta^{(0)}_{22} (\mu_L)\right]_{\rm LR} &=&
                        \eta_4^{-24/25} \eta_5^{-24/23},   \\
\left[\eta^{(1)}_{11}(\mu_L)\right]_{\rm LR}&=&0.9279\,{\eta_{4}}^{-24/25 }\,{\eta_{5}}^{-24/23 } - 0.0029\,{\eta_{4}}^{{28/25}}\,{\eta_{5}}^{-24/23 }\\
\nonumber&&+{\eta_{4}}^{{3/25}}\,{\eta_{5}}^{{3/23}}\,\left( -2.0241 - 0.0753\,\eta_{4} + 1.1744\,\eta_{4}\,\eta_{5} \right), \\
\left[\eta^{(1)}_{12}(\mu_L)\right]_{\rm LR}&=&-1.3918\,{\eta_{4}}^{-24/25 }\,{\eta_{5}}^{-24/23 } +
0.0043\,{\eta_{4}}^{{28/25}}\,{\eta_{5}}^{-24/23 }\\
\nonumber&& + 1.3875\,{\eta_{4}}^{{28/25}}\,{\eta_{5}}^{{26/23}},\\
\left[\eta^{(1)}_{21}(\mu_L)\right]_{\rm LR}&=&-0.0019\,{\eta_{4}}^{{28/25}}\,{\eta_{5}}^{-24/23 } + 5.0000\,{\eta_{4}}^{{1/25}}\,{\eta_{5}}^{{3/23}}\\
\nonumber&&+{\eta_{4}}^{{3/25}}\,{\eta_{5}}^{{3/23}}\,\left( -16.6828 - 0.0502\,\eta_{4} + 0.7829\,\eta_{4}\,\eta_{5} \right) \\
\nonumber&&+{\eta_{4}}^{-24/25 }\,{\eta_{5}}^{-24/23 }\,\left( -4.4701 - 0.8327\,\eta_{4} + 16.2548\,\eta_{4}\,\eta_{5} \right), \\
\left[\eta^{(1)}_{22}(\mu_L)\right]_{\rm LR}&=&0.0029\,{\eta_{4}}^{{28/25}}\,{\eta_{5}}^{-24/23 } + 0.9250\,{\eta_{4}}^{{28/25}}\,{\eta_{5}}^{{26/23}}\\
\nonumber&&+{\eta_{4}}^{-24/25 }\,{\eta_{5}}^{-24/23 }\,\left( 6.7052 + 1.2491\,\eta_{4} - 8.8822\,\eta_{4}\,\eta_{5} \right).
\eea
{\underline{SLL-Sector}}
\bea
\left[\eta^{(0)}_{11} (\mu_L)\right]_{\rm SLL} &=&
                        1.0153 \eta_4^{-0.5810} \eta_5^{-0.6315}
                        - 0.0153 \eta_4^{0.6610} \eta_5^{0.7184}    ,\\
\left[\eta^{(0)}_{12} (\mu_L)\right]_{\rm SLL} &=&
                        1.9325 (\eta_4^{-0.5810} \eta_5^{-0.6315}
                        - \eta_4^{0.6610} \eta_5^{0.7184})    ,\\
\left[\eta^{(0)}_{21} (\mu_L)\right]_{\rm SLL} &=&
                        0.0081 (\eta_4^{0.6610} \eta_5^{0.7184}
                        - \eta_4^{-0.5810} \eta_5^{-0.6315})   ,\\
\left[\eta^{(0)}_{22} (\mu_L)\right]_{\rm SLL} &=&
                        1.0153 \eta_4^{0.6610} \eta_5^{0.7184}
                        - 0.0153 \eta_4^{-0.5810} \eta_5^{-0.6315}  ,\\
\left[\eta^{(1)}_{11}(\mu_L)\right]_{\rm SLL}&=&
0.0020\,{\eta_{4}}^{1.6610}\,{\eta_{5}}^{-0.6315} - 0.0334\,{\eta_{4}}^{0.4190}\,{\eta_{5}}^{0.7184}\\
\nonumber&&+{\eta_{4}}^{-0.5810}\,{\eta_{5}}^{-0.6315}\,\left( 4.2458 + 0.5700\,\eta_{4} - 5.2272\,\eta_{4}\,\eta_{5} \right) \\
\nonumber&&+{\eta_{4}}^{0.6610}\,{\eta_{5}}^{0.7184}\,\left( 0.3640 + 0.0064\,\eta_{4} + 0.0724\,\eta_{4}\,\eta_{5} \right) ,\\
\left[\eta^{(1)}_{12}(\mu_L)\right]_{\rm SLL}&=&
0.0038\,{\eta_{4}}^{1.6610}\,{\eta_{5}}^{-0.6315} - 4.2075\,{\eta_{4}}^{0.4190}\,{\eta_{5}}^{0.7184}\\
\nonumber&&+{\eta_{4}}^{-0.5810}\,{\eta_{5}}^{-0.6315}\,\left( 8.0810 + 1.0848\,\eta_{4} - 38.8778\,\eta_{4}\,\eta_{5} \right) \\
\nonumber&&+{\eta_{4}}^{0.6610}\,{\eta_{5}}^{0.7184}\,\left( 45.9008 + 0.8087\,\eta_{4} - 12.7939\,\eta_{4}\,\eta_{5} \right) ,\\
\left[\eta^{(1)}_{21}(\mu_L)\right]_{\rm SLL}&=&
-0.0011\,{\eta_{4}}^{1.6610}\,{\eta_{5}}^{-0.6315} + 0.0003\,{\eta_{4}}^{0.4190}\,{\eta_{5}}^{0.7184}\\
\nonumber&&+{\eta_{4}}^{0.6610}\,{\eta_{5}}^{0.7184}\,\left( -0.0534 - 0.0034\,\eta_{4} - 0.0380\,\eta_{4}\,\eta_{5} \right) \\
\nonumber&&+{\eta_{4}}^{-0.5810}\,{\eta_{5}}^{-0.6315}\,\left( 0.0587 - 0.0045\,\eta_{4} + 0.0415\,\eta_{4}\,\eta_{5} \right) ,\\
\left[\eta^{(1)}_{22}(\mu_L)\right]_{\rm SLL}&=&
-0.0020\,{\eta_{4}}^{1.6610}\,{\eta_{5}}^{-0.6315} + 0.0334\,{\eta_{4}}^{0.4190}\,{\eta_{5}}^{0.7184}\\
\nonumber&&+{\eta_{4}}^{-0.5810}\,{\eta_{5}}^{-0.6315}\,\left( 0.1117 - 0.0086\,\eta_{4} + 0.3083\,\eta_{4}\,\eta_{5} \right) \\
\nonumber&&+{\eta_{4}}^{0.6610}\,{\eta_{5}}^{0.7184}\,\left( -6.7398 - 0.4249\,\eta_{4} + 6.7219\,\eta_{4}\,\eta_{5} \right).
\eea

\subsection{$\eta$-Factors for $K^0-\overline{K}^0$ mixing with
  $\mu_K={\cal O}(1{\rm GeV})$}

The formulae for the QCD factors $\hat{\eta}(\mu_K, \mu_t) \equiv
\hat{\eta}(\mu_K)$ relating the coefficients $C_i(\mu_K)$ and $C_i(\mu_t)$
are rather long and will not be presented. These factors can be obtained
using the relation
\begin{equation}        \label{composite_ev}
        \hat{\eta}(\mu_K) = \hat{\rho}(\mu_K) \hat{\eta}(\mu_L)
\end{equation}
with $\hat{\eta}(\mu_L)$ given in section~\ref{sec_eta_at_muL} and
$\hat{\rho}(\mu_K)$ given below. When calculating~(\ref{composite_ev}),
terms of ${\cal O}(\alpha_s^2)$ should be removed.

\newsection{$\rho$-Factors in the NDR Scheme}
\label{sec:rho}
These factors allow to calculate $\langle Q_i(\mu_L)\rangle$ from 
$\langle Q_i(\mu_K)\rangle$ with $\mu_K<\mu_c$. They can be expressed in
terms of 
\be
\label{eTa34}
\eta_3\equiv\frac{\alpha_s^{(3)}(\mu_c)}{\alpha_s^{(3)}(\mu_K)}
\qquad{\rm and}\qquad
\tilde\eta_4\equiv\frac{\alpha_s^{(4)}(\mu_L)}{\alpha_s^{(4)}(\mu_c)}.
\ee
{\underline{VLL-Sector}}
\bea
\left[\rho^{(0)} (\mu_K)\right]_{\rm VLL} &=&
                        \eta_3^{2/9} \tilde\eta_4^{6/25}    ,\\
\left[\rho^{(1)} (\mu_K)\right]_{\rm VLL} &=&
1.8951 \eta_3^{2/9} \tilde\eta_4^{6/25} - 0.1033 \eta_3^{11/9} \tilde\eta_4^{6/25}
-1.7917 \eta_3^{11/9} \tilde\eta_4^{31/25}.
\eea
{\underline{LR-Sector}}
\bea
\left[\rho^{(0)}_{11} (\mu_K)\right]_{\rm LR} &=&
                        \eta_3^{1/9} \tilde\eta_4^{3/25}    ,\\
\left[\rho^{(0)}_{12} (\mu_K)\right]_{\rm LR} &=& 0   ,\\
\left[\rho^{(0)}_{21} (\mu_K)\right]_{\rm LR} &=&
                        \frac23 (\eta_3^{1/9} \tilde\eta_4^{3/25} -
                                \eta_3^{-8/9} \tilde\eta_4^{-24/25} )    ,\\
\left[\rho^{(0)}_{22} (\mu_K)\right]_{\rm LR} &=&
                        \eta_3^{-8/9} \tilde\eta_4^{-24/25}   ,\\
\left[\rho^{(1)}_{11}(\mu_K)\right]_{\rm LR}&=&
0.9306\,{\eta_{3}}^{-{8/9} }\,{\tilde\eta_4}^{-{24/25} } - 0.0027\,{\eta_{3}}^{{10/9}}\,{\tilde\eta_4}^{-{24/25} }\\
\nonumber&&+{\eta_{3}}^{{1/9}}\,{\tilde\eta_4}^{{3/25}}\,\left( -1.9784 - 0.0457\,\eta_{3} + 1.0962\,\eta_{3}\,\tilde\eta_4 \right) ,\\
\left[\rho^{(1)}_{12}(\mu_K)\right]_{\rm LR}&=&
-1.3958\,{\eta_{3}}^{-{8/9} }\,{\tilde\eta_4}^{-{24/25} } +0.0040\,{\eta_{3}}^{{10/9}}\,{\tilde\eta_4}^{-{24/25} }\\
\nonumber&& + 1.3918\,{\eta_{3}}^{{10/9}}\,{\tilde\eta_4}^{{28/25}},\\
\label{rho21LR}
\left[\rho^{(1)}_{21}(\mu_K)\right]_{\rm LR}&=&
-3.8570\,{\eta_{3}}^{-{8/9} }\,{\tilde\eta_4}^{-{24/25} }\\
\nonumber&& + {\eta_{3}}^{{1/9}}\,{\tilde\eta_4}^{-{24/25} }\,\left( -0.6113 - 0.0018\,\eta_{3} + 20.4220\,\tilde\eta_4 \right) \\
\nonumber&&+{\eta_{3}}^{{1/9}}\,{\tilde\eta_4}^{{3/25}}\,\left( -16.6523 -
  0.7407\,\log (\eta_{3}) - 0.0305\,\eta_{3} +
  0.7308\,\eta_{3}\,\tilde\eta_4 \right) ,\\
\left[\rho^{(1)}_{22}(\mu_K)\right]_{\rm LR}&=&
5.7855\,{\eta_{3}}^{-{8/9} }\,{\tilde\eta_4}^{-{24/25} }+0.9279\,{\eta_{3}}^{{10/9}}\,{\tilde\eta_4}^{{28/25}} \\
\nonumber&&+ {\eta_{3}}^{{1/9}}\,{\tilde\eta_4}^{-{24/25} }\,\left( 0.9170 + 0.0027\,\eta_{3} - 7.6331\,\tilde\eta_4 \right).\eea
{\underline{SLL-Sector}}
\bea
\left[\rho^{(0)}_{11} (\mu_K)\right]_{\rm SLL} &=&
                        1.0153 \eta_3^{-0.5379} \tilde\eta_4^{-0.5810} -
                        0.0153 \eta_3^{0.6120} \tilde\eta_4^{0.6610} ,\\
\left[\rho^{(0)}_{12} (\mu_K)\right]_{\rm SLL} &=&
                        1.9325 ( \eta_3^{-0.5379} \tilde\eta_4^{-0.5810} -
                                \eta_3^{0.6120} \tilde\eta_4^{0.6610} )   ,\\
\left[\rho^{(0)}_{21} (\mu_K)\right]_{\rm SLL} &=&
                        0.0081 ( \eta_3^{0.6120} \tilde\eta_4^{0.6610} -
                                \eta_3^{-0.5379} \tilde\eta_4^{-0.5810} )  ,\\
\left[\rho^{(0)}_{22} (\mu_K)\right]_{\rm SLL} &=&
                        1.0153 \eta_3^{0.6120} \tilde\eta_4^{0.6610} -
                                0.0153 \eta_3^{-0.5379}
                                \tilde\eta_4^{-0.5810}  ,\\
\left[\rho^{(1)}_{11}(\mu_K)\right]_{\rm SLL}&=&
0.0019\,{\eta_{3}}^{1.6120}\,{\tilde\eta_4}^{-0.5810} - 0.0663\,{\eta_{3}}^{0.4621}\,{\tilde\eta_4}^{0.6610}\\
\nonumber&&+{\eta_{3}}^{-0.5379}\,{\tilde\eta_4}^{-0.5810}\,\left( 3.8487 + 0.3952\,\eta_{3} - 4.6906\,\eta_{3}\,\tilde\eta_4 \right) \\
\nonumber&&+{\eta_{3}}^{0.6120}\,{\tilde\eta_4}^{0.6610}\,\left( 0.4264 + 0.0039\,\eta_{3} + 0.0808\,\eta_{3}\,\tilde\eta_4 \right) ,\\
\left[\rho^{(1)}_{12}(\mu_K)\right]_{\rm SLL}&=&
0.0036\,{\eta_{3}}^{1.6120}\,{\tilde\eta_4}^{-0.5810} - 8.3647\,{\eta_{3}}^{0.4621}\,{\tilde\eta_4}^{0.6610}\\
\nonumber&&+{\eta_{3}}^{-0.5379}\,{\tilde\eta_4}^{-0.5810}\,\left( 7.3253 + 0.7521\,\eta_{3} - 42.0005\,\eta_{3}\,\tilde\eta_4 \right) \\
\nonumber&&+{\eta_{3}}^{0.6120}\,{\tilde\eta_4}^{0.6610}\,\left( 53.7722 + 0.4933\,\eta_{3} - 11.9813\,\eta_{3}\,\tilde\eta_4 \right) ,\\
\left[\rho^{(1)}_{21}(\mu_K)\right]_{\rm SLL}&=&
-0.0010\,{\eta_{3}}^{1.6120}\,{\tilde\eta_4}^{-0.5810} + 0.0005\,{\eta_{3}}^{0.4621}\,{\tilde\eta_4}^{0.6610}\\
\nonumber&&+{\eta_{3}}^{0.6120}\,{\tilde\eta_4}^{0.6610}\,\left( -0.0519 - 0.0021\,\eta_{3} - 0.0425\,\eta_{3}\,\tilde\eta_4 \right) \\
\nonumber&&+{\eta_{3}}^{-0.5379}\,{\tilde\eta_4}^{-0.5810}\,\left( 0.0628 - 0.0031\,\eta_{3} + 0.0372\,\eta_{3}\,\tilde\eta_4 \right) ,\\
\left[\rho^{(1)}_{22}(\mu_K)\right]_{\rm SLL}&=&
-0.0019\,{\eta_{3}}^{1.6120}\,{\tilde\eta_4}^{-0.5810} + 0.0663\,{\eta_{3}}^{0.4621}\,{\tilde\eta_4}^{0.6610}\\
\nonumber&&+{\eta_{3}}^{-0.5379}\,{\tilde\eta_4}^{-0.5810}\,\left( 0.1196 - 0.0060\,\eta_{3} + 0.3331\,\eta_{3}\,\tilde\eta_4 \right) \\
\nonumber&&+{\eta_{3}}^{0.6120}\,{\tilde\eta_4}^{0.6610}\,\left( -6.5470 -
  0.2592\,\eta_{3} + 6.2950\,\eta_{3}\,\tilde\eta_4 \right).
\eea

Regarding the appearance of $\log\left(\eta_3\right)$ in eq. (\ref{rho21LR})
in the LR sector, we direct the reader's attention to eq. (\ref{NLO_ev}) and
the fact that in the LR sector for $f=3$ the form of the evolution
operator given there breaks down, as the matrix $\hat J$ has a singularity
at $f=3$. However, taking the limit $f\to3$ of the complete
expression (\ref{NLO_ev}), a finite result exhibiting the aforementioned
term ${\cal O}(\alpha_s) \log\left(\eta_3\right)$ is obtained \cite{BJL}. In
accordance with the convention of (\ref{NLO_conv1}), (\ref{NLO_conv2}) we treat it as an
NLO contribution to the evolution operator.

\newsection{Numerical Results}
\label{sec:num}

\begin{figure}
\begin{center}
\epsfxsize=12cm\leavevmode\epsffile{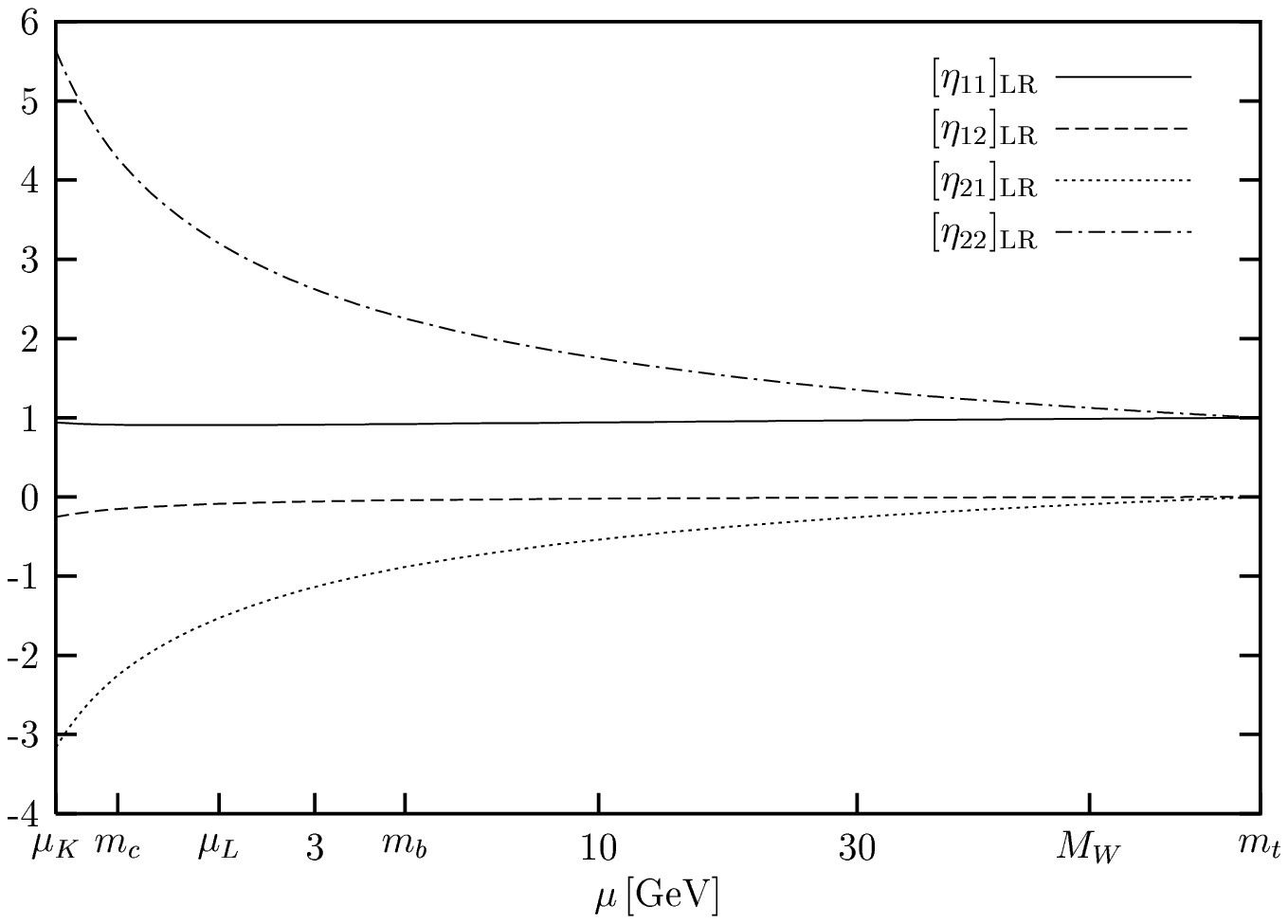}
\caption{The $\left[\eta_{ij}\right]_{\rm LR}$ factors as functions of
  $\mu$ for $\alpha_s^{(5)}(M_Z)=0.118$. \label{fig1}}
\end{center}
\end{figure}

\begin{figure}
\begin{center}
\epsfxsize=12.3cm\leavevmode\epsffile{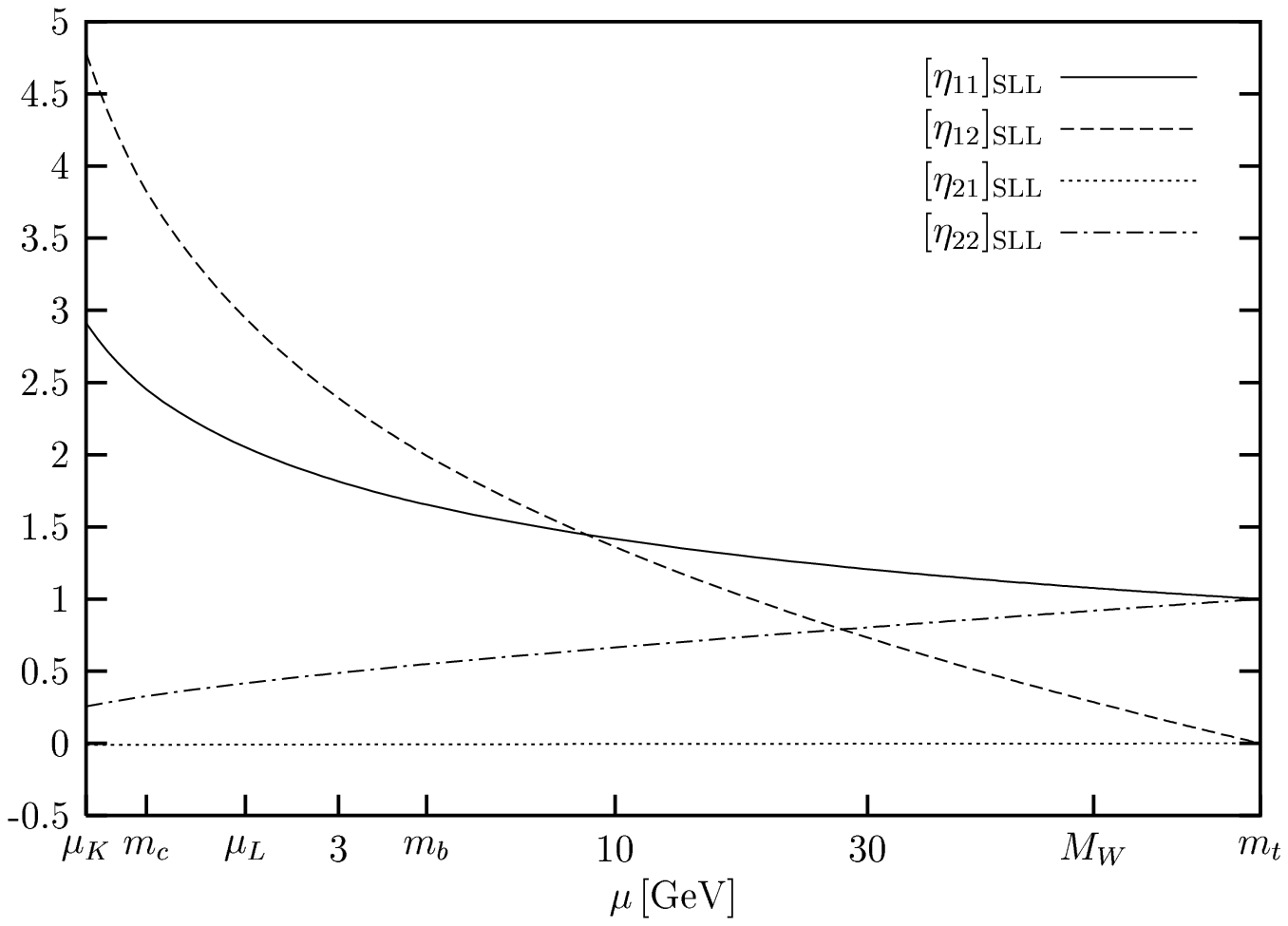}
\caption{The $\left[\eta_{ij}\right]_{\rm SLL}$ factors as functions of
  $\mu$ for $\alpha_s^{(5)}(M_Z)=0.118$\label{fig2}}
\end{center}
\end{figure}

In tables \ref{table1}--\ref{table4} we give the numerical values for 
the $\eta_{ij}$ and $\rho_{ij}$ factors in the NDR scheme. To this end we have used 
$\alpha_s^{(5)}(M_Z)=0.118\pm0.003$ with the corresponding values of
$\alpha_s^{(f)}$
and $\Lambda^{(f)}_{\overline{MS}}$ for $f=4$ and $f=3$ theories, see
table 2 in the first paper in \cite{AJBLH}.
Moreover we have set $\mu_t=m_t(m_t)= 166$ GeV,
 $\mu_b=4.4$ GeV, $\mu_L=2.0$ GeV, $\mu_c=1.3$ GeV and $\mu_K=1.0$ GeV.
In order to illustrate the
effect of the NLO corrections we show also the results in the LO. 
In doing this we have, however, used
the two-loop expression for $\alpha_s$ in both the LO and the NLO
parts. In figs. \ref{fig1} and \ref{fig2} we show the factors
$\left[\eta_{ij}(\mu)\right]_{\rm LR}$ and
$\left[\eta_{ij}(\mu)\right]_{\rm SLL}$ versus $\mu$ setting
$\alpha_s^{(5)}(M_Z)=0.118$. 

Let us recall that in the absence of renormalization group effects
$\left[\eta(\mu)\right]_{\rm VLL} = \left[\rho(\mu)\right]_{\rm VLL} = 1$ and
$\left[\hat\eta(\mu)\right]_a$ and $\left[\hat\rho(\mu)\right]_a$ 
are unit matrices. Renormalization group effects generate non-diagonal
elements in these matrices and renormalize
$\left[\eta(\mu)\right]_{\rm VLL}$ and the diagonal elements in
$\left[\hat\eta(\mu)\right]_a$ and $\left[\hat\rho(\mu)\right]_a$ away
from unity. 

Inspecting tables \ref{table1}--\ref{table4} and
figs. \ref{fig1}--\ref{fig2} we observe the following pattern:
\begin{itemize}
\item Large renormalization group effects are found for the diagonal
  entries $\left[\eta_{22}(\mu)\right]_{\rm LR}$ and
  $\left[\eta_{11}(\mu)\right]_{\rm SLL}$ which for
  $\alpha_s^{(5)}(M_Z)=0.118$ and $\mu=\mu_K=1$ GeV are enhanced by
  factors of 5.6 and 2.9, respectively. On the other hand
  $\left[\eta_{22}(\mu)\right]_{\rm SLL}$ is strongly suppressed down
  to 0.26 at $\mu_K=1$ GeV.
\item Similarly the renormalization group effects in the non-diagonal
  entries $\left[\eta_{21}(\mu)\right]_{\rm LR}$ and
  $\left[\eta_{12}(\mu)\right]_{\rm SLL}$ are large. For $\mu_K=1.0$
  GeV and $\alpha_s^{(5)}(M_Z)=0.118$ they become as large as $-3.2$
  and 4.8, respectively. This implies that $C_2^{\rm LR}(\mu)$ is
  strongly affected by the presence of the operator $Q_1^{\rm
    LR}$. Similarly $C_1^{\rm SLL}(\mu)$ is strongly affected by the
  presence of  $Q_2^{\rm SLL}$.
\item These enhancements and suppressions are more pronounced after
  the inclusion of NLO corrections. The largest NLO corrections, in
  the ball park of $25\%$, are found for the elements
  $\left[\eta_{21}(\mu_K)\right]_{\rm LR}$,
  $\left[\eta_{22}(\mu_K)\right]_{\rm LR}$ and
  $\left[\eta_{22}(\mu_K)\right]_{\rm SLL}$.
\item  $\left[\eta_{11}(\mu)\right]_{\rm LR}$ and $\eta_{\rm
    VLL}(\mu)$ are both suppressed but this suppression is at most by
  $10\%$ and $30\%$, respectively.
\item $\left[\eta_{12}(\mu)\right]_{\rm LR}$ is small   and
  $\left[\eta_{21}(\mu)\right]_{\rm SLL}$ negligible in the full range
  of $\mu$ considered. This implies that $C_1^{\rm LR}(\mu)$ and
  $C_2^{\rm SLL}(\mu)$ are essentially unaffected by the presence of
  the operators $Q_2^{\rm LR}$ and $Q_1^{\rm SLL}$, respectively.
\item Similar patterns are observed for the
  $\left[\rho_{ij}(\mu)\right]_a$ factors.
\end{itemize}
On the basis of this pattern we conclude that the renormalization
group effects strongly enhance the Wilson coefficients $C_2^{\rm LR}$
and $C_1^{\rm SLL}$ and strongly suppress $C_2^{\rm SLL}$ with respect
to their values at $\mu_t$. The
corresponding effects in $C_1^{\rm VLL}$ and $C_1^{\rm LR}$ are
substantially smaller.

\begin{table}
\begin{center}
\begin{tabular}{|l||c|c||c|c||c|c|}
\hline
\TT&\multicolumn{2}{c||}{$\alpha_s^{(5)} (M_Z) =0.115$}&
       \multicolumn{2}{c||}{$\alpha_s^{(5)} (M_Z) =0.118$}&
           \multicolumn{2}{c|}{$\alpha_s^{(5)} (M_Z) =0.121$}\\
\hline
\TT  & LO  & NLO & LO  & NLO & LO  & NLO  \\
\hline
$\TT\left[\eta (\mu_b)\right]_{\rm VLL}$  &0.835&0.847&0.829&0.842&0.823&0.836\\[2mm]
$\left[\eta_{11} (\mu_b)\right]_{\rm LR}$ &0.914&0.923&0.911&0.921&0.907&0.919\\[2mm]
$\left[\eta_{12} (\mu_b)\right]_{\rm LR}$ &0&$-$0.037&0&$-$0.041&0&$-$0.045\\[2mm]
$\left[\eta_{21} (\mu_b)\right]_{\rm LR}$ &$-$0.760&$-$0.835&$-$0.801&$-$0.885&$-$0.845&$-$0.939\\[2mm]
$\left[\eta_{22} (\mu_b)\right]_{\rm LR}$ &2.054&2.181&2.112&2.254&2.176&2.334\\[2mm]
$\left[\eta_{11} (\mu_b)\right]_{\rm SLL}$&1.560&1.621&1.587&1.654&1.616&1.690\\[2mm]
$\left[\eta_{12} (\mu_b)\right]_{\rm SLL}$&1.809&1.910&1.884&1.993&1.962&2.082\\[2mm]
$\left[\eta_{21} (\mu_b)\right]_{\rm SLL}$&$-$0.008&$-$0.006&$-$0.008&$-$0.007&$-$0.008&$-$0.007\\[2mm]
$\left[\eta_{22} (\mu_b)\right]_{\rm SLL}$&0.595&0.563&0.583&0.549&0.570&0.535 \\[2mm]
\hline
\end{tabular}
\caption{Numerical values for the $\eta$-factors for
  $B^0_{d,s}-\overline{B}^0_{d,s}$ mixing.\label{table1}}
\end{center}
\end{table}

\begin{table}
\begin{center}
\begin{tabular}{|l||c|c||c|c||c|c|}
\hline
\TT&\multicolumn{2}{c||}{$\alpha_s^{(5)} (M_Z) =0.115$}&
       \multicolumn{2}{c||}{$\alpha_s^{(5)} (M_Z) =0.118$}&
           \multicolumn{2}{c|}{$\alpha_s^{(5)} (M_Z) =0.121$}\\
\hline
\TT  & LO  & NLO & LO  & NLO & LO  & NLO  \\
\hline
\TT$\left[\eta (\mu_L)\right]_{\rm VLL}$     &0.778&0.796&0.768&0.788&0.757&0.780\\[2mm]
$\left[\eta_{11} (\mu_L)\right]_{\rm LR}$ &0.882&0.906&0.876&0.906&0.870&0.906\\[2mm]
$\left[\eta_{12} (\mu_L)\right]_{\rm LR}$ &0&$-$0.076&0&$-$0.087&0&$-$0.101\\[2mm]
$\left[\eta_{21} (\mu_L)\right]_{\rm LR}$ &$-$1.236&$-$1.401&$-$1.336&$-$1.530&$-$1.449&$-$1.677\\[2mm]
$\left[\eta_{22} (\mu_L)\right]_{\rm LR}$ &2.735&3.009&2.879&3.200&3.043&3.420\\[2mm]
$\left[\eta_{11} (\mu_L)\right]_{\rm SLL}$&1.859&1.976&1.918&2.052&1.984&2.138\\[2mm]
$\left[\eta_{12} (\mu_L)\right]_{\rm SLL}$&2.586&2.775&2.732&2.946&2.892&3.137\\[2mm]
$\left[\eta_{21} (\mu_L)\right]_{\rm SLL}$&$-$0.011&$-$0.009&$-$0.011&$-$0.009&$-$0.012&$-$0.009\\[2mm]
$\left[\eta_{22} (\mu_L)\right]_{\rm SLL}$&0.480&0.438&0.461&0.417&0.442&0.394\\[2mm]
\hline
\end{tabular}
\caption{Numerical values for the $\eta$-factors for $K^0-\overline{K}^0$ mixing with
  $\mu=\mu_L=2$ GeV.\label{table2}}
\end{center}
\end{table}

\begin{table}
\begin{center}
\begin{tabular}{|l||c|c||c|c||c|c|}
\hline
\TT&\multicolumn{2}{c||}{$\alpha_s^{(5)} (M_Z) =0.115$}&
       \multicolumn{2}{c||}{$\alpha_s^{(5)} (M_Z) =0.118$}&
           \multicolumn{2}{c|}{$\alpha_s^{(5)} (M_Z) =0.121$}\\
\hline
\TT   & LO  & NLO & LO  & NLO & LO  & NLO  \\
\hline
\TT$\left[\eta (\mu_K)\right]_{\rm VLL}$  &0.701& 0.735& 0.681& 0.720& 0.658& 0.705\\[2mm]
$\left[\eta_{11} (\mu_K)\right]_{\rm LR}$ &0.837& 0.921& 0.825& 0.941& 0.811& 0.978\\[2mm]
$\left[\eta_{12} (\mu_K)\right]_{\rm LR}$ &0 & $-$0.194& 0 & $-$0.254& 0 & $-$0.347\\[2mm]
$\left[\eta_{21} (\mu_K)\right]_{\rm LR}$ &$-$2.199& $-$2.657& $-$2.545& $-$3.159& $-$3.006& $-$3.861\\[2mm]
$\left[\eta_{22} (\mu_K)\right]_{\rm LR}$ &4.136& 4.875& 4.643& 5.630& 5.320& 6.688\\[2mm]
$\left[\eta_{11} (\mu_K)\right]_{\rm SLL}$&2.392& 2.663& 2.566& 2.912& 2.787& 3.243\\[2mm]
$\left[\eta_{12} (\mu_K)\right]_{\rm SLL}$&3.836& 4.273& 4.223& 4.782& 4.702& 5.442\\[2mm]
$\left[\eta_{21} (\mu_K)\right]_{\rm SLL}$&$-$0.016& $-$0.011& $-$0.018& $-$0.012& $-$0.020& $-$0.012\\[2mm]
$\left[\eta_{22} (\mu_K)\right]_{\rm SLL}$&0.346& 0.291& 0.314& 0.255& 0.279& 0.217\\[2mm]
\hline
\end{tabular}
\caption{Numerical values for the $\eta$-factors for $K^0-\overline{K}^0$ mixing with
  $\mu_K=1$ GeV.\label{table3}}
\end{center}
\end{table}

\begin{table}
\begin{center}
\begin{tabular}{|l||c|c||c|c||c|c|}
\hline
\TT&\multicolumn{2}{c||}{$\alpha_s^{(5)} (M_Z) =0.115$}&
       \multicolumn{2}{c||}{$\alpha_s^{(5)} (M_Z) =0.118$}&
           \multicolumn{2}{c|}{$\alpha_s^{(5)} (M_Z) =0.121$}\\
\hline
\TT   & LO  & NLO & LO  & NLO & LO  & NLO  \\
\hline
\TT$\left[\rho (\mu_K)\right]_{\rm VLL}$     &0.902&0.923&0.887&0.914&0.870&0.905\\[2mm]
$\left[\rho_{11} (\mu_K)\right]_{\rm LR}$ &0.950&0.955&0.942&0.951&0.933&0.947\\[2mm]
$\left[\rho_{12} (\mu_K)\right]_{\rm LR}$ &0&$-$0.045&0&$-$0.060&0&$-$0.083\\[2mm]
$\left[\rho_{21} (\mu_K)\right]_{\rm LR}$ &$-$0.375&$-$0.448&$-$0.447&$-$0.548&$-$0.544&$-$0.688\\[2mm]
$\left[\rho_{22} (\mu_K)\right]_{\rm LR}$ &1.512&1.620&1.613&1.762&1.748&1.963\\[2mm]
$\left[\rho_{11} (\mu_K)\right]_{\rm SLL}$&1.293&1.357&1.345&1.431&1.413&1.533\\[2mm]
$\left[\rho_{12} (\mu_K)\right]_{\rm SLL}$&1.029&1.176&1.190&1.381&1.394&1.650\\[2mm]
$\left[\rho_{21} (\mu_K)\right]_{\rm SLL}$&$-$0.004&$-$0.003&$-$0.005&$-$0.003&$-$0.006&$-$0.003\\[2mm]
$\left[\rho_{22} (\mu_K)\right]_{\rm SLL}$&0.744&0.688&0.710&0.641&0.670&0.584\\[2mm]
\hline
\end{tabular}
\caption{Numerical values for the $\rho$-factors with $\mu_K=1$ GeV
  and $\mu_L=2$ GeV.\label{table4}}
\end{center}
\end{table}

\clearpage
\newsection{General Remarks}
\subsection{Renormalization Scheme Dependence}
The evolution matrix $\hat U$ is renormalization scheme dependent. It is instructive to recall
how this scheme dependence is canceled in physical amplitudes by
considering a single operator $Q$. Then the $\Delta F=2$ amplitude
reads
\be
A(\Delta F=2)=
\langle H_{\rm eff}\rangle = \frac{G_F^2}{16\pi^2}M_W^2  V_{\rm CKM} 
C(\mu)\langle Q(\mu) \rangle~.
\ee
The Wilson coefficient is given by
\begin{equation}\label{CC}
  C(\mu) =  U(\mu,\mu_t)  C(\mu_t)   
\end{equation}
where
\begin{equation}\label{UMNLO}
 U (\mu,\mu_t) = \Biggl\lbrack 1 + {{\alpha_s (\mu)}\over{4\pi}} J
\Biggl\rbrack \Biggl\lbrack {{\alpha_s (\mu_t)}\over{\alpha_s (\mu)}}
\Biggl\rbrack^P \Biggl\lbrack 1 - {{\alpha_s (\mu_t)}\over{4\pi}} J
\Biggl\rbrack 
\end{equation}
with
\begin{equation}
P = {{\gamma^{(0)}}\over{2\beta_0}},~~~~~~~~~ J = {{P}\over{\beta_0}}
\beta_1 - {{\gamma^{(1)}}\over{2\beta_0}}
\end{equation} 
and
\be\label{init}
C(\mu_t)=C_0+\frac{\alpha_s(\mu_t)}{4\pi} C_1.
\ee
$C_0$ and $C_1$ depend generally on $m_t$, $M_W$ and the masses
of new particles in extensions of the SM. 

Now, the renormalization scheme dependence of $C_1$ is canceled
by the one of $J$ in the last square bracket in (\ref{UMNLO}). The
scheme dependence of $J$ in the first square bracket in (\ref{UMNLO})
is canceled by the scheme dependence of $\langle Q(\mu)\rangle$. The
power $P$ and the coefficient $C_0$ are scheme independent.

\subsection{Transformation to Different Renormalization Schemes}
Once the Wilson coefficients $C_i(\mu)$ 
have been calculated in the NDR scheme, 
they can be transformed to a different renormalization scheme $A$
by means of
\be
\vec C^{\rm A}(\mu)=
\left(1-\frac{\alpha_s(\mu)}{4\pi}\Delta\hat r^T_{\rm NDR\to A}\right)
\vec C^{\rm NDR}(\mu)~.
\ee
Likewise the matrix elements $\langle Q_i(\mu)\rangle$ can be
transformed from scheme A to the NDR scheme:
\be
\langle\vec Q(\mu)\rangle_{\rm NDR}=
\left(1+\frac{\alpha_s(\mu)}{4\pi}\Delta\hat r_{\rm A\to NDR}\right)
\langle\vec Q(\mu)\rangle_{\rm A}~,
\qquad \Delta\hat r_{\rm A\to NDR}=-\Delta \hat r_{\rm NDR \to A}~.
\ee
The transformation matrices $\Delta\hat r_{\rm NDR \to RI}$ from the NDR scheme to 
the RI schemes of \cite{CET0} can be found in section 5 of \cite{BMU}.

\subsection{$\eta_B$ and $\eta_2$ Factors in the SM}
At this point we would like to warn the reader that the QCD factors $\eta_B=0.55$ 
\cite{BJW90,UKJS} and $\eta_2$ \cite{BJW90} used in the analysis
of $B^0_{d,s}-\overline{B}^0_{d,s}$ mixing and of the top quark 
contribution to $\varepsilon_K$ in the SM should not be identified
with the factors $\left[\eta (\mu_b)\right]_{\rm VLL}$ and
$\left[\eta (\mu_K)\right]_{\rm VLL}$ presented in this paper.

The factors $\eta_B$  and $\eta_2$  are discussed in detail in 
\cite{BBL,AJBLH,BJW90}. See in particular the expressions 
(12.10) and
(13.3) in \cite{BBL} for $\eta_2$ and $\eta_B$, respectively. Using
these expressions it is straightforward to find the relation between 
$\eta_B$ and $\left[\eta (\mu_b)\right]_{\rm VLL}$. It reads
\be
\label{rel}
\left[\eta (\mu_b)\right]_{\rm VLL} C^{\rm VLL}_{\rm SM}(\mu_t)=
\left[\alpha_s^{(5)}(\mu_b)\right]^{-6/23}
\Biggl\lbrack 1 + {{\alpha_s^{(5)}(\mu_b)}\over{4\pi}} J_5
\Biggl\rbrack \eta_B 4 S_0(x_t)
\ee
where $C^{\rm VLL}_{\rm SM}(\mu_t)$ includes NLO corrections
calculated in \cite{BJW90,UKJS}, $J_5=1.627$ in the NDR scheme and 
$4 S_0(x_t)$ with $x_t=m_t^2(\mu_t)/M_W^2$ is the LO expression for
$C^{\rm VLL}_{\rm SM}(\mu_t)$ that is  
obtained from box diagrams with top quark exchanges without QCD corrections. $\eta_B$ 
in contrast to $\left[\eta (\mu_b)\right]_{\rm VLL}$ is renormalization
 scheme independent and does not depend on $\mu_b$. The latter 
dependence has been factored out as seen on the r.h.s of (\ref{rel}).
Notice that the QCD corrections to $C^{\rm VLL}_{\rm SM}(\mu_t)$ have
been absorbed into $\eta_B$. An analogous relation between 
$\eta_2$ and $\left[\eta (\mu_K)\right]_{\rm VLL}$ can be found.

\subsection{Going Beyond the SM}
In the SM
\be
\label{eff_ham_BBbar}
\langle \bar B^{0}|H_{\rm eff}^{\Delta B=2}|B^0\rangle_{\rm SM} =
\frac{G_F^2}{48 \pi^2} M_W^2 m_B F_B^2 \hat B_B (V_{tb}^* V_{td})^2
\eta_B 4 S_0 (x_t)
\ee
where $\hat B_B$ is the renormalization group invariant parameter
defined by
\be
\hat B_B = B_1^{\rm VLL}(\mu_b)
\left[\alpha_s^{(5)}(\mu_b)\right]^{-6/23} \left[1+\frac{\alpha_s^{(5)}(\mu_b)}{4
    \pi} J_5\right],
\ee
with $B_1^{\rm VLL}(\mu)$ defined in (\ref{matelm1}).
$F_B$ is the $B$-meson decay constant and $\eta_B$ is the QCD factor
defined in (\ref{rel}).

In the extensions of the SM with minimal flavour violation (MFV) and without
contributions from new operators it is useful to generalize
(\ref{eff_ham_BBbar}) to
\be
\label{eff_ham_susy}
\langle \bar B^{0}|H_{\rm eff}^{\Delta B=2}|B^0\rangle =
\frac{G_F^2}{48 \pi^2} M_W^2 m_B F_B^2 \hat B_B (V_{tb}^* V_{td})^2
\eta_B 4 F_{tt}
\ee
where
\be
F_{tt} = S_0(x_t) + \frac{\eta_B^{\rm new}}{\eta_B} S_0^{\rm new} .
\ee
Here $\eta_B^{\rm new}$ and $S_0^{\rm new}$ describe new physics
contributions in analogy to $\eta_B$ and $S_0(x_t)$, respectively.
That is 
\be
\left[\eta (\mu_b)\right]_{\rm VLL} C_{\rm new}^{\rm VLL}(\mu_t)=
\left[\alpha_s^{(5)}(\mu_b)\right]^{-6/23}
\Biggl\lbrack 1 + {{\alpha_s^{(5)}(\mu_b)}\over{4\pi}} J_5
\Biggl\rbrack \eta^{\rm new}_B 4 S^{\rm  new}_0(x_t),
\ee
where $C_{\rm new}^{\rm VLL} (\mu_t)$ is the new physics contribution
to $C_1^{\rm VLL}(\mu_t)$ and $4 S_0^{\rm new}$ results from new
physics contributions without QCD corrections.

In more complicated models in which new flavour-violating couplings are
present and the full set of operators (\ref{normal})
is relevant, it appears to be most useful to evaluate new
physics contributions using simply 
\be
\langle \bar B^{0}|H_{\rm eff}^{\Delta B=2}|B^0\rangle_{\rm new} =
\frac{G_F^2}{16 \pi^2} M_W^2 \sum C_i (\mu) \langle \bar B^{0}|Q_i(\mu)|B^0\rangle
\ee
with $C_i(\mu)$ evaluated by means of the formulae in sections
\ref{sec:basic} and \ref{sec:eta}.
Similar comments apply to the $K^0-\bar K^0$ system with obvious
replacements.
\newsection{Phenomenological Applications}
\subsection{Hadronic Matrix Elements for $K^0-\bar{K}^0$ Mixing}

The matrix elements $\langle \bar{K}^0|Q_i(\mu)| K^0\rangle \equiv \langle
Q_i(\mu)\rangle$ contributing to $K^0-\bar{K}^0$
mixing can be written as
\begin{eqnarray}
\label{matelm1}
        \langle Q_1^{\rm VLL}(\mu) \rangle &=&
                        \f{1}{3} m_K F_K^2 B_1^{\rm VLL} (\mu) ,\\
        \langle Q_1^{\rm LR}(\mu) \rangle &=&
                        -\f{1}{6} R(\mu)\, m_K F_K^2 B_1^{\rm LR} (\mu) ,\\
        \langle Q_2^{\rm LR}(\mu) \rangle &=&
                        \f{1}{4} R(\mu)\, m_K F_K^2 B_2^{\rm LR} (\mu) ,\\
        \langle Q_1^{\rm SLL}(\mu) \rangle &=&
                        -\f{5}{24} R(\mu)\, m_K F_K^2 B_1^{\rm SLL}
                        (\mu) ,\\
\label{matelm5}
        \langle Q_2^{\rm SLL}(\mu) \rangle &=&
                        -\f{1}{2} R(\mu)\, m_K F_K^2 B_2^{\rm SLL} (\mu),
\end{eqnarray}
where
\be
        R(\mu) = \left( \f{m_K}{m_s(\mu) + m_d(\mu)} \right)^2
\ee
and $F_K$ is the $K$-meson decay constant. Let us calculate the  non-perturbative parameters
$B^a_i(\mu)$ using the lattice results of~\cite{latt_lit}
discussed in~\cite{CET}. In the Landau RI scheme (LRI) the $B^a_i(\mu)$ factors
are given by
\begin{eqnarray}
\nnb
        \left[ B_1^{\rm VLL} (\mu) \right]_{\rm LRI} &=&
                \left[ B_1 (\mu) \right]_{\rm LRI}      ,\\
\label{b1vll}
        \left[ B_1^{\rm LR} (\mu) \right]_{\rm LRI} &=&
                \left[ B_5 (\mu) \right]_{\rm LRI}      ,\hspace{2.5cm}
        \left[ B_2^{\rm LR} (\mu) \right]_{\rm LRI} =
                \left[ B_4 (\mu) \right]_{\rm LRI}      ,\\
\nnb
        \left[ B_1^{\rm SLL} (\mu) \right]_{\rm LRI} &=&
                \left[ B_2 (\mu) \right]_{\rm LRI}      ,\hspace{2.4cm}
        \left[ B_2^{\rm SLL} (\mu) \right]_{\rm LRI} =
                \left[ \f{5}{3} B_2 (\mu) - \f{2}{3} B_3(\mu) \right]_{\rm LRI},
\end{eqnarray}
where $B_i(\mu),\;\; i=1,\ldots,5$ are the non-perturbative factors entering
the matrix elements in the operator basis of~\cite{CET}.

In order to find the matrix elements~(\ref{matelm1})--(\ref{matelm5}) in the NDR scheme we use the
results of~\cite{BMU}, which allow us to relate the $B_i$ factors in the LRI
and NDR schemes. We find
\begin{eqnarray}
\label{B1VLL}
        \left[ B_1^{\rm VLL} (\mu) \right]_{\rm NDR} &=&
                \left[ B_1^{\rm VLL} (\mu) \right]_{\rm LRI} +
                \frac{\alpha_s^{(4)}(\mu)}{4 \pi} r^{\rm VLL}
                        \left[ B_1^{\rm VLL} (\mu) \right]_{\rm LRI}    ,\\
        \left[ B_1^{\rm LR} (\mu) \right]_{\rm NDR} &=&
                \left[ B_1^{\rm LR} (\mu) \right]_{\rm LRI} +
                \frac{\alpha_s^{(4)}(\mu)}{4 \pi}
                        \left[ r_{11}^{\rm LR} B_1^{\rm LR} (\mu)
                                - \f{3}{2} r_{12}^{\rm LR} B_2^{\rm LR} (\mu)
                        \right]_{\rm LRI}                               ,\\
        \left[ B_2^{\rm LR} (\mu) \right]_{\rm NDR} &=&
                \left[ B_2^{\rm LR} (\mu) \right]_{\rm LRI} +
                \frac{\alpha_s^{(4)}(\mu)}{4 \pi}
                        \left[ -\f{2}{3} r_{21}^{\rm LR} B_1^{\rm LR} (\mu)
                                +r_{22}^{\rm LR} B_2^{\rm LR} (\mu)
                        \right]_{\rm LRI}                               ,\\
        \left[ B_1^{\rm SLL} (\mu) \right]_{\rm NDR} &=&
                \left[ B_1^{\rm SLL} (\mu) \right]_{\rm LRI} +
                \frac{\alpha_s^{(4)}(\mu)}{4 \pi}
                        \left[ r_{11}^{\rm SLL} B_1^{\rm SLL} (\mu)
                                + \f{12}{5} r_{12}^{\rm SLL} B_2^{\rm SLL}(\mu)
                        \right]_{\rm LRI},\\
\label{B2SLL}
        \left[ B_2^{\rm SLL} (\mu) \right]_{\rm NDR} &=&
                \left[ B_2^{\rm SLL} (\mu) \right]_{\rm LRI} +
                \frac{\alpha_s^{(4)}(\mu)}{4 \pi}
                        \left[ \f{5}{12} r_{21}^{\rm SLL} B_1^{\rm SLL} (\mu)
                                +r_{22}^{\rm SLL} B_2^{\rm SLL} (\mu)
                        \right]_{\rm LRI},
\end{eqnarray}
where
\bea
 r^{\rm VLL} &\equiv&\Delta r^{\rm VLL}_{\rm LRI\to NDR} =  0.8785, \\
 \hat{r}^{\rm LR} &\equiv& \Delta r^{\rm LR}_{\rm LRI\to NDR} =
                      \left( \matrix{-1.1288 & -6.7726 \cr
                                     0.3069  & 10.8712} \right),\\
 \hat{r}^{\rm SLL} &\equiv& \Delta r^{\rm SLL}_{\rm LRI\to NDR} = 
                      \left( \matrix{ 5.6438 & 0.2140 \cr
                                     12.9387 & 2.6892} \right).
\eea
Now, the $B_i$ factors presented in~\cite{CET,latt_lit} read for $\mu=\mu_L=2~{\rm GeV}$
as follows:
\begin{eqnarray}
\nonumber
        \left[ B_1 \right]_{\rm LRI} &=& 0.60 \pm 0.06, \hspace{3cm}
        \left[ B_2 \right]_{\rm LRI} = 0.66 \pm 0.04    ,\\
        \left[ B_3 \right]_{\rm LRI} &=& 1.05 \pm 0.12, \hspace{3cm}
        \left[ B_4 \right]_{\rm LRI} = 1.03 \pm 0.06    ,\\
\nonumber
        \left[ B_5 \right]_{\rm LRI} &=& 0.73 \pm 0.10.
\end{eqnarray}
Using the central values for these parameters, we find by means of
(\ref{b1vll})
\begin{eqnarray}
\nonumber
        \left[ B_1^{\rm VLL} \right]_{\rm LRI} &=& 0.60  ,\\
        \left[ B_1^{\rm LR} \right]_{\rm LRI} &=& 0.73, \hspace{4cm}
        \left[ B_2^{\rm LR} \right]_{\rm LRI} = 1.03  ,\\
\nonumber
        \left[ B_1^{\rm SLL} \right]_{\rm LRI} &=& 0.66, \hspace{4cm}
        \left[ B_2^{\rm SLL} \right]_{\rm LRI} = 0.40.
\end{eqnarray}
Finally, setting $\alpha_s^{(4)}(2~{\rm GeV}) = 0.306$ and using
(\ref{B1VLL})--(\ref{B2SLL})
we obtain in the NDR scheme for $\mu = \mu_L = 2~{\rm GeV}$
\begin{eqnarray}
\nonumber
        \left[ B_1^{\rm VLL} \right]_{\rm NDR} &=& 0.61, \\
\label{blabla}
        \left[ B_1^{\rm LR} \right]_{\rm NDR} &=& 0.96, \hspace{4cm}
        \left[ B_2^{\rm LR} \right]_{\rm NDR} = 1.30, \\
\nonumber
        \left[ B_1^{\rm SLL} \right]_{\rm NDR} &=& 0.76, \hspace{4cm}
        \left[ B_2^{\rm SLL} \right]_{\rm NDR} = 0.51.
\end{eqnarray}
We observe that the scheme dependence in the LR and SLL sectors is large,
amounting to a shift of the $B_i$ factors by roughly 30\% and 20\%,
respectively. The corresponding scheme dependence in the VLL sector amounts
to 2\%.

Setting $F_K = 160~{\rm MeV}$ and $m_K = 498~{\rm MeV}$ we obtain at $\mu = 2~{\rm GeV}$
\begin{eqnarray}
        \langle Q_1^{\rm VLL} \rangle_{\rm NDR} &=&
                        0.26 \cdot 10^{-2} {\rm GeV}^3          ,\\
        \langle Q_1^{\rm LR} \rangle_{\rm NDR} &=&
                -3.83 \left[
                        \f{115 {\rm MeV}}{m_s(2 {\rm GeV}) + m_d(2 {\rm GeV})}
                      \right]^2 \cdot 10^{-2} {\rm GeV}^3       ,\\
        \langle Q_2^{\rm LR} \rangle_{\rm NDR} &=&
                7.77 \left[
                        \f{115 {\rm MeV}}{m_s(2 {\rm GeV}) + m_d(2 {\rm GeV})}
                      \right]^2 \cdot 10^{-2} {\rm GeV}^3       ,\\
        \langle Q_1^{\rm SLL} \rangle_{\rm NDR} &=&
                -3.79 \left[
                        \f{115 {\rm MeV}}{m_s(2 {\rm GeV}) + m_d(2 {\rm GeV})}
                      \right]^2 \cdot 10^{-2} {\rm GeV}^3       ,\\
        \langle Q_2^{\rm SLL} \rangle_{\rm NDR} &=&
                -6.10 \left[
                        \f{115 {\rm MeV}}{m_s(2 {\rm GeV}) + m_d(2 {\rm GeV})}
                      \right]^2 \cdot 10^{-2} {\rm GeV}^3.
\end{eqnarray}
\subsection{$\Delta M_K$, $\Delta M_B$ and $\varepsilon_K$}

Next we would like to present general formulae for the mass
differences $\Delta M_K$ and $\Delta M_B$ in the $K^0-\bar K^0$
and $B^0-\bar B^0$ systems  and for the CP-violating
parameter $\varepsilon_K$. In the case of $\Delta M_K$ and 
$\varepsilon_K$ our formulae are valid for the contributions
of heavy internal particles with masses higher than $M_W$. The
known SM contributions from internal charm quark exchanges and
 mixed charm-top exchanges \cite{HN} have to be added separately.

We have
\be
\Delta M_K= 
2 {\rm Re}\langle \bar K^0| H^{\Delta S=2}_{\rm eff}|K^0\rangle~,
\ee

\be
\Delta M_B=
2  |\langle \bar B^0| H^{\Delta B=2}_{\rm eff}|B^0\rangle|~,
\ee
\be
\varepsilon_K=\frac{\exp(i\pi/4)}{\sqrt{2}\Delta M_K}
  {\rm Im}\langle \bar K^0| H^{\Delta S=2}_{\rm eff}|K^0\rangle~.
\ee
The matrix element $\langle\bar K^0|H_{\rm eff}^{\Delta
  S=2}|K^0\rangle$ can be written as follows

\begin{eqnarray}\label{hds2}
\langle \bar K^0| H^{\Delta S=2}_{\rm eff}|K^0\rangle &=&
\frac{G_F^2}{48\pi^2}M_W^2 m_K F_K^2 
\left[ P_1^{\rm VLL} (C^{\rm VLL}_1(\mu_t)+C^{\rm VRR}_1(\mu_t))\right.
\nonumber\\
& & + P_1^{\rm LR} C^{\rm LR}_1(\mu_t)     
 +P_2^{\rm LR} C^{\rm LR}_2(\mu_t) \nonumber\\
& & \left.+
P_1^{\rm SLL} (C^{\rm SLL}_1(\mu_t)+C^{\rm SRR}_1(\mu_t))
+P_2^{\rm SLL} (C^{\rm SLL}_2(\mu_t)+C^{\rm SRR}_2(\mu_t))\right]
\end{eqnarray}
with
\be
\label{p1_factor}
P_1^{\rm VLL}=\left[\eta (\mu_L)\right]_{\rm VLL} 
B^{\rm VLL}_1(\mu_L),
\ee
\be
P_1^{\rm LR} =-\frac{1}{2} \left[\eta_{11} (\mu_L)\right]_{\rm LR}
               \left[B^{\rm LR}_1(\mu_L)\right]_{\rm eff}
+\frac{3}{4} \left[\eta_{21} (\mu_L)\right]_{\rm LR}
               \left[B^{\rm LR}_2(\mu_L)\right]_{\rm eff},
\ee
\be
P_2^{\rm LR}=-\frac{1}{2} \left[\eta_{12} (\mu_L)\right]_{\rm LR}
               \left[B^{\rm LR}_1(\mu_L)\right]_{\rm eff}
+\frac{3}{4} \left[\eta_{22} (\mu_L)\right]_{\rm LR}
               \left[B^{\rm LR}_2(\mu_L)\right]_{\rm eff},
\ee

\be
P_1^{\rm SLL} =-\frac{5}{8} \left[\eta_{11} (\mu_L)\right]_{\rm SLL}
               \left[B^{\rm SLL}_1(\mu_L)\right]_{\rm eff}
-\frac{3}{2} \left[\eta_{21} (\mu_L)\right]_{\rm SLL}
               \left[B^{\rm SLL}_2(\mu_L)\right]_{\rm eff},
\ee
\be
P_2^{\rm SLL}=-\frac{5}{8} \left[\eta_{12} (\mu_L)\right]_{\rm SLL}
               \left[B^{\rm SLL}_1(\mu_L)\right]_{\rm eff}
-\frac{3}{2} \left[\eta_{22} (\mu_L)\right]_{\rm SLL}
               \left[B^{\rm SLL}_2(\mu_L)\right]_{\rm eff}.
\ee

In the case of the SM and MFV models one can use (\ref{eff_ham_BBbar})
and (\ref{eff_ham_susy}) instead of (\ref{p1_factor}).
In writing these formulae we have absorbed the CKM factors into
$C_i(\mu_t)$. The effective parameters
$\left[B^a_i(\mu_L)\right]_{\rm eff}$ are defined by

\be
\label{Balat}
\left[B^a_i(\mu_L)\right]_{\rm eff}\equiv
\left(\frac{m_K}{m_s(\mu_L)+m_d(\mu_L)}\right)^2 B^a_i(\mu_L)=
18.75\left[\frac{115~{\rm MeV}}{m_s(\mu_L)+m_d(\mu_L)}\right]^2
B^a_i(\mu_L).
\ee
In the case of $B^0-\bar B^0$ mixing one has to make the replacements
$\mu_L\to \mu_b$ and $m_K F_K^2\to m_B F_B^2$. Then in the case of
$B_d^0-\bar B^0_d$ system
\be
\left[B^a_i(\mu_b)\right]_{\rm eff}\equiv
\left(\frac{m_B}{m_b(\mu_b)+m_d(\mu_b)}\right)^2 B^a_i(\mu_b)=
1.44\left[\frac{4.4~{\rm GeV}}{m_b(\mu_b)+m_d(\mu_b)}\right]^2
B^a_i(\mu_b),
\ee
with an analogous formula for the $B_s^0-\bar B^0_s$ system.

We would like to emphasize that these formulae together with
the QCD factors $\eta_{ij}$ presented in Section 3 are valid
for any extension of the SM. In particular the coefficients 
$P^a_i$ are universal. New physics contributions enter only
the coefficients $C^a_i(\mu_t)$. The latter have to be evaluated
in the NDR scheme in order to cancel the scheme dependence of
the universal coefficients $P^a_i$. In the process of multiplying
$C^a_i(\mu_t)$ and $P^a_i$ terms ${\cal O}(\alpha_s^2)$ have to
be removed.

It is instructive to evaluate the coefficients $P^a_i$ for the
$K^0-\bar K^0$ system. Setting $\mu_L=2$ GeV, 
$\Lambda_{\overline{MS}}^{(4)}=325$ MeV, $m_s(\mu_L)+m_d(\mu_L)=115$ MeV 
and using the values for $B^a_i$ in (\ref{blabla}) we find
\be
P_1^{\rm VLL}=0.48,
\ee
\be
P_1^{\rm LR} =-36.1, \qquad P_2^{\rm LR}= 59.3,
\ee
\be
P_1^{\rm SLL} =-18.1, \qquad P_2^{\rm SLL}=-32.2.
\ee
We observe that the coefficients $P_i^{\rm LR}$ and $P_i^{\rm SLL}$
are by two orders of magnitude larger than $P_1^{\rm VLL}$. This 
originates in the strong enhancement of the QCD factors $\eta_{ij}$ 
for the LR and SLL (SRR) operators and in the chiral enhancement of their
matrix elements as seen in (\ref{Balat}). Consequently even small new physics 
contributions to $C^{\rm LR}_i(\mu_t)$ and $C^{\rm SLL}_i(\mu_t)$ 
can play an important role in the phenomenology \cite{{BMZ91},{CET}}.

In the case of $B^0-\bar B^0$ mixing the chiral enhancement of the
hadronic matrix elements in the LR and SLL sectors is absent. 
Moreover, the QCD factors $\eta_{ij}$ are smaller than in the 
$K^0-\bar K^0$ mixing. Consequently the coefficients 
$P_i^{\rm LR}$ and $P_i^{\rm SLL}$ are smaller in this case.
As lattice results are not yet available for all the relevant hadronic 
matrix elements in the $B$ system \cite{LL} we will
set $B_i^a(m_b)=1$. Taking $m_B=5.28$ GeV, $\mu_b=4.4$ GeV,  
$m_b(\mu_b)+m_d(\mu_b)=4.4$ GeV and 
$\Lambda_{\overline{MS}}^{(5)}=226$ MeV we find
\be
P_1^{\rm VLL}=0.84,
\ee
\be
P_1^{\rm LR} =-1.62, \qquad P_2^{\rm LR}= 2.46,
\ee
\be
P_1^{\rm SLL} =-1.47, \qquad P_2^{\rm SLL}=-2.98.
\ee

\newsection{Summary}
\label{sec:sum}
We have presented analytic formulae for the QCD renormalization group factors
relating the Wilson coefficients $C_i(\mu_t)$ and $C_i(\mu)$, with
$\mu_t={\cal O}(m_t)$ and $\mu <\mu_t$, of the $\Delta F=2$ dimension 
six four-quark operators $Q_i$. 
The formulae presented in section 3 are given in the NDR scheme but
are otherwise universal and apply to
the Standard Model and all its possible extensions.
In order to complete the evaluation of $\Delta F=2$ amplitudes, the 
QCD factors presented here have to be combined with the Wilson 
coefficients $C_i(\mu_t)$ evaluated in a given model at the short 
distance scale $\mu_t$ and with the hadronic matrix elements 
$\langle Q_i(\mu)\rangle$ evaluated at $\mu=\mu_b$, $\mu=\mu_L$ or 
$\mu=\mu_K$ dependently on the process considered.
$C_i(\mu_t)$ and $\langle Q_i(\mu)\rangle$ have to be computed in 
the NDR scheme in order to obtain a renormalization scheme independent 
answer for the physical amplitudes. 

We have also presented analytic
formulae for the QCD factors relating the matrix elements 
$\langle Q_i (2~{\rm GeV})\rangle$ and $\langle Q_i (\mu_K)\rangle$ 
with $\mu_K<2$ GeV. These formulae allow
the comparison of the matrix elements obtained in lattice simulations with 
those obtained in approaches which use lower renormalization scales. 

Our numerical analysis shows that the renormalization-group effects are
very large in the LR and SLL sectors. This in particular is the case for
the elements $[\eta_{21}(\mu)]_{\rm LR}$, $[\eta_{22}(\mu)]_{\rm LR}$,
$[\eta_{11}(\mu)]_{\rm SLL}$, $[\eta_{12}(\mu)]_{\rm SLL}$ and
$[\eta_{22}(\mu)]_{\rm SLL}$ and is in
accordance with the previous analyses~\cite{BMZ91,CET}.
The NLO corrections amount typically to 5-15\% except for the elements
$[\eta_{21}(\mu)]_{\rm LR}$, $[\eta_{22}(\mu)]_{\rm LR}$ and
$[\eta_{22}(\mu)]_{\rm SLL}$, where in the
case of $\mu=1 {\rm GeV}$ they can reach 25\%. As a result of this
pattern the renormalization group effects strongly enhance 
the Wilson coefficients $C_2^{\rm LR}$
and $C_1^{\rm SLL}$ and strongly suppress $C_2^{\rm SLL}$ with respect
to their values at $\mu_t$. The
corresponding effects in $C_1^{\rm VLL}$ and $C_1^{\rm LR}$ are
substantially smaller.

Finally we have presented expressions for the mass
differences $\Delta M_K$ and $\Delta M_B$ and the CP-violating parameter
$\epsilon_K$ in terms of the non-perturbative parameters $B^a_i$ and the
Wilson coefficients $C_i(\mu_t)$. These formulae include
renormalization group effects at the NLO level and allow to calculate
straightforwardly $\Delta M_K$, $\Delta M_B$ and $\epsilon_K$ in any
extension of the SM once the Wilson coefficients $C_i(\mu_t)$ and the
non-perturbative parameters $B_i^a$ are known in the NDR scheme. In
the case of $K^0-\bar K^0$ mixing we have presented the results for
$\left[B_i^a(2\,{\rm GeV})\right]_{\rm NDR}$ using the lattice results
obtained in the LRI scheme. The corresponding results for the $B$
system should be available this year.

\ \\
{\Large\bf Acknowledgements}

We would like to thank Christoph Bobeth and Janusz Rosiek for critical
reading of the manuscript and enlightening discussions.

This work has been supported in part by the German Bundesministerium
f\"ur Bildung und Forschung under the contract 05HT9WOA0.

\ \\
\appendix

\section*{Appendix}
\section{One-Loop and Two-Loop Anomalous Dimension Matrices}   \label{app_adm}

We give below the one-loop and two-loop anomalous dimension matrices. The two-loop 
expressions are given in the NDR scheme \cite{BMU}.
\bea
\begin{array}{ll}
\gamma^{(0)\rm VLL} = 4, &
\gamma^{(1)\rm VLL} = -7 + \f{4}{9} f,\\[4mm]
\hat{\gamma}^{(0)\rm LR} = \left( \begin{array}{ccc} 
         2 &~& 12 \\[1mm]
         0 && -16 \end{array} \right), &
\hat{\gamma}^{(1)\rm LR} = \left( \begin{array}{ccc} 
         \f{71}{3} - \f{22}{9} f   &~~&    198 - \f{44}{3} f \\[1mm] 
         \f{225}{4} - 2 f          &&    -\f{1343}{6}  + \f{68}{9} f
         \end{array} \right),\\[8mm]
\hat{\gamma}^{(0)\rm SLL} = \left( \begin{array}{ccc}
        -10  &~& \f{1}{6} \\[1mm]
        -40  && \f{34}{3}
\end{array} \right),\qquad&
\hat{\gamma}^{(1)\rm SLL} = \left( \begin{array}{ccc}
        -\f{1459}{9} + \f{74}{9} f      &~~&    -\f{35}{36} - \f{1}{54} f  \\[1mm]
        -\f{6332}{9} + \f{584}{9} f     &&      \f{2065}{9} -
        \f{394}{27} f
\end{array} \right).
\end{array}
\eea

\section{Expansion of the evolution matrices $\hat U$ in $\alpha_s$}
\label{app_higher_orders}
Recall the renormalization group equation to which~(\ref{UM}) is the formal
solution. At NLO it reads, written in terms of $\alpha_s$,
\begin{equation}        \label{nlo_rge}
        \f{d \hU(\mu_1, \mu_2)}{d \alpha_s} = \left[
                -\f{\hat{\gamma}^{(0)T}}{2 \beta_0} \f{1}{\alpha_s}
                +\left( \f{\beta_1}{2 \beta_0^2}\hat{\gamma}^{(0)T}
                        - \f{1}{2 \beta_0}\hat{\gamma}^{(1)T}
                \right) \frac{1}{4 \pi}
        \right] \hU(\mu_1, \mu_2).
\end{equation}
At the leading order, where only the term $\propto 1/\alpha_s$ is
kept,~(\ref{nlo_rge}) has the (exact) solution
\begin{equation}
        \hU^{(0)}(\mu_1, \mu_2) =
                \left(\frac{\alpha_s(\mu_2)}{\alpha_s(\mu_1)}\right)^
                        {\frac{\hat{\gamma}^{(0)T}}{2 \beta_0}}
        = \exp \left(\frac{\hat{\gamma}^{(0)T}}{2 \beta_0}
                                \log \frac{\alpha_s(\mu_2)}{\alpha_s(\mu_1)}
                \right).
\end{equation}
At the next-to-leading order one has~\cite{BBL}
\begin{equation}        \label{NLO_ev}
        \hU(\mu_1, \mu_2) = \left( {\hat{1}} + \frac{\alpha^{(f)}_s(\mu_1)}{4 \pi} \hJ_f \right)
                \hU_{f}^{(0)} \left( {\hat{1}} - \frac{\alpha^{(f)}_s(\mu_2)}{4 \pi} \hJ_f
        \right),
\end{equation}
where higher orders in the parentheses have been omitted. The algorithm
for constructing the matrix $\hJ_f$ can be found in~\cite{{BBL},{AJBLH},{BJL}}.
Eq.~(\ref{NLO_ev}) holds in a theory with a given number $f$ of active
quark flavours.
When evolving across a quark threshold, as for instance in~(\ref{UUU1}),
one finds
\mathindent0pt
\begin{eqnarray}
        \hU(\mu_L, \mu_t) \!\!\! &=& \!\!\!
            \left(\hat{1} + \frac{\alpha^{(4)}_s(\mu_L)}{4 \pi} \hJ_4 \right)
            \hU^{(0)}_{f=4} \left(\hat{1} - \frac{\alpha^{(4)}_s(\mu_b)}{4 \pi} \hJ_4
                \right)
            \left( \hat{1} + \frac{\alpha^{(5)}_s(\mu_b)}{4 \pi} \hJ_5 \right)
            \hU^{(0)}_{f=5} \left(\hat{1} - \frac{\alpha^{(5)}_s(\mu_t)}{4 \pi} \hJ_5
                \right).         \nonumber \\
        & & \label{NLO_2step}
\end{eqnarray}
\mathindent1cm
In light of the fact that ${\cal O}(\alpha_s^2)$ and higher terms have been
dropped in~(\ref{nlo_rge}) and~(\ref{NLO_ev}), we adopt the convention
\begin{eqnarray}
\label{NLO_conv1}
        \eta^{\rm P}
        &=& \left(\frac{\alpha_s(\mu_2)}{\alpha_s(\mu_1)} \right)^
                {\rm P}
        = {\cal O} (\alpha_s^0) ,\\
\label{NLO_conv2}
        \log(\eta) &=& {\cal O}(\alpha_s^0)
\end{eqnarray}
and drop all ${\cal O}(\alpha_s^2)$ and higher terms in~(\ref{NLO_2step})
and similar expressions. For the desired two- and three-step evolution
matrices, one obtains
\mathindent0pt
\begin{eqnarray}
        \hU(\mu_L, \mu_t) &=&
                \hU^{(0)}_{f=4}(\mu_L, \mu_b) \hU^{(0)}_{f=5}(\mu_b, \mu_t)
                                                                \nonumber \\
        &  &    + \frac{\alpha^{(4)}_s(\mu_L)}{4 \pi} \left[
                        \hJ_4 \hU^{(0)}_{f=4} \hU^{(0)}_{f=5}
                        +\eta_4 \hU^{(0)}_{f=4}\left(\hJ_5-\hJ_4\right)
                                \hU^{(0)}_{f=5}
                        - \eta_4 \eta_5 \hU^{(0)}_{f=4} \hU^{(0)}_{f=5} \hJ_5
                \right],
\end{eqnarray}
\begin{eqnarray}
        \hU(\mu_K, \mu_L) &=&
                \hU^{(0)}_{f=3}(\mu_K, \mu_c) \hU^{(0)}_{f=4}(\mu_c, \mu_L)
                                                                \nonumber \\
        &  &    + \frac{\alpha^{(3)}_s(\mu_K)}{4 \pi} \left[
                        \hJ_3 \hU^{(0)}_{f=3} \hU^{(0)}_{f=4}
                        +\eta_3 \hU^{(0)}_{f=3}\left(\hJ_4-\hJ_3\right)
                                \hU^{(0)}_{f=4}
                        - \eta_3 \eta_4 \hU^{(0)}_{f=3} \hU^{(0)}_{f=4} \hJ_4
                \right],
\end{eqnarray}
\begin{eqnarray}
        \hU(\mu_K, \mu_t) &=&
                \hU^{(0)}_{f=3}(\mu_K, \mu_c) \hU^{(0)}_{f=4}(\mu_c, \mu_b)
                        \hU^{(0)}_{f=5}(\mu_b, \mu_t)           \nonumber \\
        &  &    + \frac{\alpha^{(3)}_s(\mu_K)}{4 \pi} \left[
                        \hJ_3 \hU^{(0)}_{f=3} \hU^{(0)}_{f=4} \hU^{(0)}_{f=5}
                        + \eta_3 \hU^{(0)}_{f=3}\left(\hJ_4-\hJ_3\right)
                                \hU^{(0)}_{f=4} \hU^{(0)}_{f=5} \right.
                                                                \nonumber\\
        & &     \left.  + \eta_3 \eta_4 \hU^{(0)}_{f=3} \hU^{(0)}_{f=4}
                                \left(\hJ_5-\hJ_4\right) \hU^{(0)}_{f=5}
                        - \eta_3 \eta_4 \eta_5 \hU^{(0)}_{f=3}
                                \hU^{(0)}_{f=4} \hU^{(0)}_{f=5} J_5
                \right],
\end{eqnarray}
\mathindent1cm
where we have suppressed some obvious arguments in the LO evolution
matrices~$U^{(0)}$ in order not to unnecessarily clutter the expressions.
\section{The Evolution Matrix $\hat U(\mu_t,\mu_s)$}
For completeness we give here the elements of the evolution matrix
$\hat U(\mu_t, \mu_s)$ in a $f=6$ flavour theory with
$\mu_s>\mu_t$. The renormalization group evolution from $\mu_s$ down to
$\mu_t$ can even be included as in (\ref{two_step_ev}) when $\mu_s$ is only by a
factor of two higher than $m_t$. However, it is only necessary when
$\mu_s>4 m_t$ in order to avoid large logarithms.

The formulae given below are not as general as the
ones given in section~\ref{sec:eta}. They apply only to the evolution of new
physics contributions which do not involve SM particles except for the
number of quark flavours entering $\alpha_s$ and the anomalous
dimensions of the operators (\ref{normal}). This is for instance the case
considered in \cite{{BMZ91},{CET}} in which squarks and gluinos have
been integrated out at a scale $\mu_s\gg\mu_t$. On the other hand the
renormalization group analysis of charged Higgs contributions with
$M_{H^\pm}\gg m_t$ would be more complicated as both $H^{\pm}$ and top
can be simultaneously exchanged in box diagrams. Integrating out first
$H^{\pm}$ and subsequently the top would introduce bilocal structures
for $\mu_t<\mu<\mu_s$ quite analogous to the study of charm
contributions to $K^0-\overline K^0$ mixing \cite{{Gilman:1983ap},{HN}}. We find then\\
{\underline{VLL-Sector}}
\bea
\left[\eta^{(0)} (\mu_t)\right]_{\rm VLL} &=& \eta_6^{6/21},    \\
\left[\eta^{(1)} (\mu_t)\right]_{\rm VLL} &=& 1.3707 (1-\eta_6) \eta_6^{6/21}.
\eea
{\underline{LR-Sector}}
\bea
\left[\eta^{(0)}_{11} (\mu_t)\right]_{\rm LR} &=& \eta_6^{3/21},   \\
\left[\eta^{(0)}_{12} (\mu_t)\right]_{\rm LR} &=& 0,   \\
\left[\eta^{(0)}_{21} (\mu_t)\right]_{\rm LR} &=&
                        \frac23(\eta_6^{3/21} -\eta_6^{-24/21}),   \\
\left[\eta^{(0)}_{22} (\mu_t)\right]_{\rm LR} &=& \eta_6^{-24/21},\\
\left[\eta^{(1)}_{11}(\mu_t)\right]_{\rm LR}&=&0.9219\,{\eta_{6}}^{-24/21 } + {\eta_{6}}^{3/21}\,\left( -2.2194 + 1.2975\,\eta_{6} \right), \\
\left[\eta^{(1)}_{12}(\mu_t)\right]_{\rm LR}&=&1.3828\,({\eta_{6}}^{24/21}-\eta_6^{-24/21}),\\
\left[\eta^{(1)}_{21}(\mu_t)\right]_{\rm LR}&=&{\eta_{6}}^{3/21}\,\left( -10.1463 + 0.8650\,\eta_{6} \right)  + {\eta_{6}}^{-24/21 }\,\left( -6.4603 + 15.7415\,\eta_{6} \right), \\
\left[\eta^{(1)}_{22}(\mu_t)\right]_{\rm LR}&=&0.9219\,{\eta_{6}}^{24/21} + {\eta_{6}}^{-24/21 }\,\left( 9.6904 - 10.6122\,\eta_{6} \right).
\eea
{\underline{SLL-Sector}}
\bea
\left[\eta^{(0)}_{11} (\mu_t)\right]_{\rm SLL} &=&
                        1.0153 \eta_6^{-0.6916} - 0.0153 \eta_6^{0.7869},  \\
\left[\eta^{(0)}_{12} (\mu_t)\right]_{\rm SLL} &=&
                        1.9325 (\eta_6^{-0.6916} - \eta_6^{0.7869}), \\
\left[\eta^{(0)}_{21} (\mu_t)\right]_{\rm SLL} &=&
                        0.0081 (\eta_6^{0.7869} - \eta_6^{-0.6916}),    \\
\left[\eta^{(0)}_{22} (\mu_t)\right]_{\rm SLL} &=&
                        1.0153 \eta_6^{0.7869} - 0.0153 \eta_6^{-0.6916},    \\
\left[\eta^{(1)}_{11}(\mu_t)\right]_{\rm SLL}&=&{\eta_{6}}^{-0.6916}\,\left( 5.6478 - 6.0350\,\eta_{6} \right)  + {\eta_{6}}^{0.7869}\,\left( 0.3272 + 0.0600\,\eta_{6} \right), \\
\left[\eta^{(1)}_{12}(\mu_t)\right]_{\rm SLL}&=&{\eta_{6}}^{-0.6916}\,\left( 10.7494 - 37.9209\,\eta_{6} \right)  + {\eta_{6}}^{0.7869}\,\left( 41.2556 - 14.0841\,\eta_{6} \right), \\
\left[\eta^{(1)}_{21}(\mu_t)\right]_{\rm SLL}&=&{\eta_{6}}^{0.7869}\,\left( -0.0618 - 0.0315\,\eta_{6} \right)  + {\eta_{6}}^{-0.6916}\,\left( 0.0454 + 0.0479\,\eta_{6} \right), \\
\left[\eta^{(1)}_{22}(\mu_t)\right]_{\rm SLL}&=&{\eta_{6}}^{-0.6916}\,\left( 0.0865 + 0.3007\,\eta_{6} \right)  + {\eta_{6}}^{0.7869}\,\left( -7.7870 + 7.3999\,\eta_{6} \right).
\eea
Here $\mu_t={\cal O}(m_t)$ and $\eta_6 =
\alpha_s^{(6)}(\mu_s)/\alpha_s^{(6)}(\mu_t)$. These results together
with those presented in section 3 and 4 allow to find $\hat
U(\mu,\mu_s)$ with $\mu<\mu_t$, see (\ref{two_step_ev}).

\renewcommand{\baselinestretch}{0.95}

\vfill\eject

\end{document}